\documentclass[twocolumn]{aastex631} 
\newcommand{\sna}{SN~Ia}
\newcommand{\sne}{SNe~Ia}
\newcommand{\msun}{\mbox{$\mathrm{M_{\odot}}$}}
\newcommand{\pyr}{\mbox {{\rm yr$^{-1}$}}}
\newcommand{\mch}{\mbox {$\mathrm{M_{Ch}}$}}
\newcommand{\mdot}{\mbox{$\mathrm{\dot{M}}$\ }}

\newcommand{\isotope}[2]{${}^{#1}$#2}
\newcommand{\nane}{${^{23}}${(Na,Ne)}\,}
\newcommand{\ecap}{${e}${-capture}}
\newcommand{\pcap}{${p}${-capture}}
\newcommand{\ncap}{${n}${-capture}}
\newcommand{\acap}{${\alpha}${-capture}}
\newcommand{\betad}{${\beta}${-decay}}
\newcommand{\mwd}{\mbox {{\rm M$_{\rm WD}$}}}
\newcommand{\gcc}{\mbox {{\rm g$\cdot$cm$^{-3}$}}}
\newcommand{\eexp}{\mbox {{$\bar{\eta}_{exp}$}}}

\received{}
\revised{}
\accepted{}
\submitjournal{\apj}

\shorttitle{\sne\ Simmering Phase}
\shortauthors{Piersanti et al.}

\begin{document}
\graphicspath{{figures/}}

\title{Pre-explosive accretion and simmering phases of Type Ia Supernovae}

\correspondingauthor{Luciano Piersanti}
\email{luciano.piersanti@inaf.it, eduardo.bravo@upc.es, 
oscar.straniero@inaf.it, sergio.cristallo@inaf.it, inma@ugr.es}

\author[0000-0002-8758-244X]{Luciano Piersanti}
\altaffiliation{INFN, Sezione di Perugia, via Pascoli, 06121, Perugia, Italy}
\affiliation{INAF-Osservatorio Astronomico d'Abruzzo, via Mentore Maggini, snc, I-64100, Teramo, Italy}
\nocollaboration{1}

\author[0000-0003-0894-6450]{Eduardo Bravo}
\affiliation{E.T.S. Arquitectura del Vall\'es, Universitat Polit\`ecnica de Catalunya, Carrer Pere Serra  
1-15, 08173 Sant Cugat del Vall\`es,Spain,}
\nocollaboration{1}

\author[0000-0002-5514-6125]{Oscar Straniero}
\altaffiliation{INFN, Laboratori Nazionali del Gran Sasso (LNGS), 67100 Assergi, Italy}
\affiliation{INAF-Osservatorio Astronomico d'Abruzzo, via Mentore Maggini, snc, I-64100, Teramo, Italy}
\nocollaboration{1}

\author[0000-0001-9683-9406]{Sergio Cristallo}
\altaffiliation{INFN, Sezione di Perugia, via Pascoli, 06121, Perugia, Italy}
\affiliation{INAF-Osservatorio Astronomico d'Abruzzo, via Mentore Maggini, snc, I-64100, Teramo, Italy}
\nocollaboration{1}

\author[0000-0002-3827-4731]{Inmaculada Dom\'\i nguez}
\affiliation{Universidad de Granada, E-18071 Granada, Spain}
\nocollaboration{1}

\begin{abstract}
In accreting WDs approaching the Chandrasekhar limit, hydrostatic carbon burning precedes the
dynamical breakout. During this {\it simmering} phase, \ecap s are energetically favored in the central region of the star,
while \betad~ are favored more outside, and the two zones are connected by a growing convective instability.
We analyze the interplay between weak interactions and convection, the so-called convective URCA process, during the simmering
phase of \sne~ progenitors and its effects on the physical and chemical properties at the explosion epoch. At variance
with previous studies, we find that the convective core powered by the carbon burning remains confined within the
\isotope{21}{(Ne,F)} URCA shell.
As a result, a much larger amount of carbon has to be consumed before the explosion which eventually occurs at
larger density than previously estimated.
In addition, we find that the extension of the convective core and its average neutronization depend on the the WD progenitor
initial metallicity. For the average neutronization in the convective core at the explosion epoch we obtain
${\overline{\eta}_{exp}} = (1.094\pm 0.143)\times 10^{-3} + (9.168\pm 0.677)\times 10^{-2}\times Z$.
Outside the convective core, the neutronization is instead determined by the initial amount of C+N+O in the progenitor star. Since
S, Ca, Cr and Mn, the elements usually exploited to evaluate the pre-explosive neutronization, are mainly produced outside the heavily
neutronized core, the problem of too high metallicity estimated for the progenitosr of the historical Tycho and Kepler \sne~
remains unsolved. 
\end{abstract}

 
\section{Introduction}\label{s:intro}

Type Ia Supernovae (\sne) are among the most reliable extragalactic distance indicators and, in turn, they play a fundamental role in our understanding of the Universe. Hubble diagrams based on \sne\ have shown the recent acceleration of the expansion rate of the Universe \citep{riess1998, perlmutter1999}, as well as the previous deceleration \citep{riess2004}.  While high redshift \sne\ are fundamental tools to probe dark energy \citep[see, for example,][]{betoule2014,jones2018}, those at lower redshifts constrain the Hubble-Lemaitre constant \citep[see, for example,][]{macaulay2019,hamuy2021}. 

The precision of \sne\  as distance indicators relies on empirical relations, such as, for example, the maximum-decline relation \citep{phillips1993,phillips1999}, that link their maximum absolute magnitudes with light curve shape and colour. Thanks to dedicated surveys collecting thousands of SNe Ia, it has been proved that these relations show a quite small dispersion, i.e., $\sim0.1$~mag \citep{scolnic2018}. This occurrence demonstrates that SNe Ia are a very homogeneous class of events, that is the key feature for being a reliable tool for cosmology.

On the other hand, the homogeneity of this class of supernovae is supported by their most popular model. It consists of a thermonuclear explosion of a carbon-oxygen white dwarf (CO WD) with a mass of the order of the Chandrasekhar limit  \citep{hoyle1960}. According to this scenario, the WD is destabilized by mass accretion from a close stellar companion. The supernova engine is a thermonuclear explosion that starts with the carbon fusion. As a result, the whole WD is incinerated and its ashes contain intermediate-mass elements and a substantial amount of iron-group isotopes. \sne\ are indeed the main contributors to the iron pollution of the interstellar gas \citep[see, for example,][]{matteucci1985,1993brab,timmes1995,prantzos2018}.  Among these nuclei, \isotope{56}{Ni} is mostly produced,  and it is the radioactive energy from the cascade \isotope{56}{Ni}$-$\isotope{56}{Co}$-$\isotope{56}{Fe}~ that powers \sne\  light curves (LCs). 
This simple scenario implies that the memory of the properties of the WD progenitor, such as the initial mass or the original chemical composition, are lost in the supernova outcomes, i.e., LCs and spectra. This occurrence is often referred as ``\textit{stellar amnesia}'' \citep{hoflich2006}. 

\sne~  still remain an intriguing mystery. First of all, no clear consensus exists concerning the nature of their stellar progenitors, the evolutionary paths up to the explosion and the explosion mechanisms. Moreover, several questions have been raised about the actual diversity of the \sne.  It is well known that the total amount of  \isotope{56}{Ni} determines the  maximum brightness and other features of the LCs. Therefore, the rather large range of values of the maximum brightness would imply that different amounts of \isotope{56}{Ni} should be produced by different supernovae. What determines these differences? Moreover, various features of the light curves correlate with the properties of the host galaxies 
\citep{hamuy1996, hamuy2000, gallagher2005, mannucci2005, rigault2013, sullivan2006, kang2020, hakobyan2021,kelsey2021}. These occurrences appear in contrast with the supposed stellar amnesia.

Several progenitor systems have been proposed \citep[see][for a more detailed review]{branch2017}. 
A critical property is the mass of the WD at the explosion time, which is strictly connected to the explosion mechanism. In the classical model, when the WD mass attains the Chandrasekhar limit,  
 the compressional heating of the whole star induces a central carbon ignition in highly degenerate conditions (Chandrasekhar-mass scenario). Basing on 1D model, the outward burning front initially moves with sub-sonic velocity (deflagration). However, in order to obtain a good reproduction of the observed spectral evolution, a transition to a detonation should occur at a certain point \citep[delayed detonation model,][]{khokhlov1991}. The physical mechanism that induces this transition is, however, unknown.  
Recently, explosions of sub-Chandrasekhar mass WDs, initially proposed to explain some sub-luminous events \citep{livne1990,ww94}, have been suggested to represent a large part of the normal \sne\  \citep{flors2020}. In these sub-Chandrasekhar WDs, the explosion is a consequence of He accretion onto a CO WD. It starts with a detonation of  the external He-rich layer. Then, an inward shock moves toward the center triggering a carbon detonation. At the opposite, rotating WDs with a mass exceeding the non-rotating Chandrasekhar limit have been proposed to explain a few super-luminous \sne\ that imply a large amount of \isotope{56}{Ni}\citep{howell2006,brown2014,hicken2007}. However, rotation is a very common feature for WDs belonging to close-binary systems and, in turn, it should play an important role for the majority of the \sne~, rather than for just a minority. 
 
Moreover, the nature of the companion object is still under debate: broadly speaking, it could be a normal star \citep[\textit{Single Degenerate}, or SD, scenario, ][]{whelan1973}
or another WD \citep[\textit{Double Degenerate}, or DD scenario, ][]{iben1984,web84}.  In the SD scenario the companion could have an H-rich envelope (classical SD)
or it may be an He star (He-donor channel), while in the DD scenario, the companion could be another CO WD (classical DD) or an He WD (He-donor channel) \citep{tutu1996,yl04,sy05,wang09}. Other scenarios have been proposed. Examples are the merging of a CO WD with the CO core of an AGB companion star, during a
 common envelope episode 
\citep[\textit{Core-Degenerate scenario} - ][]{ilkov2012}, or violent mergers of two white dwarfs \citep{pakmor2012,pakmor2013}. 

In principle, the composition of the progenitors at the time of explosion, specifically the C/O ratio and the degree of neutronization\footnote{As usual, for a mixture of $\mathcal{N}$ isotopes, the neutronization is defined as $\eta=\sum_{i}\frac{X_i}{A_i}\left ( A_i-2Z_i \right )$, where $X_i$, $A_i$ and $Z_i$ are the mass fraction, the atomic number and the 
charge number, respectively, and $i=$1,..., $\mathcal{N}$.}, influences the explosive nucleosynthesis \citep{hoflich1998,hamuy2000,inma2001,timmes2003,moreno2016,piersanti2017}. 
For instance, abundance ratios of intermediate-mass elements, like Si, and iron-peak elements in the material ejected by a SN  may provide some hints on the neutronization degree. 
In this context, \citet{badenes2008} \citep[see also][]{park2013} used the Mn to Cr abundance ratio to derived the average preexplosive
neutronization of the Tycho and Kepler \sne, namely: \eexp=$4.36\times 10^{-3}$ and $4.55\times 10^{-3}$, respectively.
In principle, the neutronization of a WD should depend on the progenitor metallicity.
Indeed, the composition of the C-O core of an intermediate-mass star is the result of both the H burning and the subsequent He burning. 
In the first evolutionary phase, the original CNO material is mainly converted into $^{14}$N.  
Later on, during He burning, $^{22}$Ne is produced through the chain  $^{14}$N$(\alpha,\gamma)^{18}$F$(\beta^+,\nu_e) ^{18}$O$(\alpha,\gamma)^{22}$Ne. In practice, the final abundance (by number) of $^{22}$Ne is equal to the original C+N+O abundance (by number). Since for the major constituents, i.e. $^{12}$C and $^{16}$O, the contribution to $\eta$ is 0, the neutronization in the CO core of an intermediate mass star is essentially due to the resulting $^{22}$Ne abundance and, in turn, to the original C+N+O. On this base, the initial metallicity of the Tycho and the Kepler progenitors would have been as large as $Z=3.5-4 Z_\odot$, values which are much larger than the average metallicity of the galactic disk. 
This occurrence suggests that the average neutronization of the exploding WD should increase during the accretion phase. Indeed, due to the mass deposition, the WD density increases up to a few $10^{9}$ \gcc. At such a high density \ecap s~ are energetically favored with respect to the reverse processes ($\beta$-decays)  and, for this reason, the neutronization is expected to rise up.

\citet{piro2008} and \citet{chamulak2008} investigated such a possibility, showing that the average neutronization increases during the so-called {\sl simmering phase}, 
{\it i.e.}, the hydrostatic C-burning phase that precedes the dynamical breakout.  
In particular, they found that the larger the carbon consumption during the simmering, the higher the pre-explosive neutronization. Nevertheless, for $Z> Z_\odot$, they found that the electro-capture contribution to the final \eexp\ is small compared to the  $Z$-dependent contribution, so that the high metallicity problem for the Tycho and Kepler supernova progenitors remains unsolved. 
Similar conclusion was  also derived by \citet{martinez2016} (hereinafter MR2016), while \citet{piersanti2017} (hereinafter P2017) found that the variation of the average neutronization during the accretion and 
the simmering phases increases as the initial metallicity of the progenitor increases, thus showing that the estimated \eexp\ for the Tycho and 
Kepler \sne\ are compatible with a progenitor metallicity of $Z\sim 2Z_\odot$, close to the average metallicity of the thin disk. 
\citet{schwab2017} (hereinafter S2017) suggest that these ``\textit{ differences ...
could arise due to differences in the net effect of the convective Urca process}''. Actually, from a careful reading of all these theoretical studies, it appears that in addition to the URCA process, the various computations were carried out under quite different assumptions about the treatments of relevant physical processes, such as the convective mixing, the heat and the transports, and the nuclear network.    
As a matter of fact, a self-consistent treatment of all the physical processes  operating during the last few thousand years prior to the explosion is still missing. The aim of the present work is to extend previous studies, by reviewing the treatment of all the relevant physical processes that may influence the final neutronization of the exploding WD. 
Then, in \S \ref{s:neutronization}, we critically analyze the evolution of an accreting WD before and after the development of the convective core.  
We review, in particular, the properties of the convective URCA processes, paying attention to identify all the possible URCA-pairs that may affect the thermal budget of the CO core and the extension of the convective region. Moreover, we investigate the effects of changing the numerical treatment of the convective instabilities, in particular, the  mixing scheme and the criterion adopted to identify unstable zones. Eventually, we check the completeness of the adopted nuclear network.  
Basing on this analysis, in \S \ref{s:literature} we show that all the models currently available for the \sne\ progenitors are inaccurate and their \eexp\ predictions are not reliable.
After discussing in \S \ref{s:preanalysis} the effects of each energy contribution associated with URCA processes and their connection with convection, 
in \S \ref{s:ngmodels} we present new models of accreting CO WDs 
up to the explosion, 
and in \S \ref{s:finalsetup} we review the effects of the often neglected \isotope{21}{(F,Ne)} URCA shell. 
Eventually, in \S \ref{s:newmodel} we discuss our updated predictions for the pre-explosive structures and we present the new relation between progenitor metallicity and  pre-explosive neutronization. 
Finally, in \S \ref{s:variation}
we explore the dependence of the simmering phase on the initial mass of the accreting WD and of its cooling age as well as on the adopted accretion rate. 
Conclusions follow in \S \ref{s:conclusion}.

In Appendix ~\ref{a:computed} we summarize all the models discussed in the present work and the setup adopted for their computation.

\section{The overall evolutionary scenario} \label{s:neutronization}

According to the classical Chandrasekhar-mass scenario for \sne~ progenitors, two distinct phases precede the explosion, namely:
the \textit{accretion phase} and the 
\textit{simmering phase}. In the former, owing to the inverse relationship between mass and radius of a degenerate star, the density of the accreting WD increases until conditions for carbon ignition are attained at the center. The latter is instead characterized by a thermonuclear runaway, as usual for a nuclear burning in degenerate conditions, and the consequent development of an extended convective core, whose external border moves progressively outward. 

A key process occurring since the accretion phase, is the activation of cyclic weak interactions, in particular {\ecap}s~ followed by {\betad}s. The former start at the center, when the density attains a critical value, typically, of a few $10^9$ \gcc. Then the reverse process, the \betad, may eventually close the cycle. Each cycle, which is called URCA process, usually involves a pair of isobars. At first, a \textit{mother}  isotope, $\mathrm{Is_1}$, evolves into the \textit{daughter}, $\mathrm{Is_2}$, according to the 
weak reaction $\mathrm{Is_1}(A,Z)+e^-\rightarrow \mathrm{Is_2}(A,Z-1)+\nu$. Then, the reaction $\mathrm{Is_2}(A,Z-1)\rightarrow \mathrm{Is_1}(A,Z)+e^{-}+\overline{\nu}$ transforms the daughter into the mothers nucleus. 
The density for which the \ecap~ on the mother nucleus and the \betad~ on the daughter have the same probability to occur  defines the location of an URCA shell. In practice, for each isobar pair, this threshold density ($\rho_{th}$) separates the more internal zone, where the \ecap s~ are favored , from the external one,  where \betad s~ dominate. Indeed, the energy released by an \ecap~ is:
\begin{equation}
E_{\text{ec}}=E_{M}-E_{\nu}+E_{K}
\label{e:eec}
\end{equation}
where $E_{M}=(M(Is_1)+m_e-M(Is_2))\cdot c^2$ 
is the mass excess between reactants and products,  $E_{\nu}$ is the energy carried away by neutrinos, 
and $E_{K}$ is the average thermal energy of the captured electrons\footnote{As usual, the difference of the thermal energy of the two ions is neglected.}. 
The latter is of the order of the electron Fermi energy that scales as $\rho^{\alpha}$, with $1/3 \leq\alpha\leq 2/3$. Since the first term is usually negative, the process becomes energetically feasible ($E_{\text{ec}}>0$) only if the Fermi energy is sufficiently large. 
This occurrence implies high density. On the contrary, \betad s~  are hampered at high density, because of the Pauli suppression. More outside, where the density is lower, \betad s~ can occur, while \ecap s~ are suppressed, because of the negative $E_{\text{ec}}$. In this case,  the energy released by the \betad~ is: 
\begin{equation}
E_{\beta}=E_{M}-E_{\overline{\nu}}-E_{K}
\label{e:ebeta}
\end{equation}
where $E_{M}=(M(Is_2)-m_e-M(Is_1))\cdot c^2$, and the other terms have the same meaning as in equation \ref{e:eec}. 
Close to the Urca shell, both reactions are active and the rate of energy production/subtraction is:
\begin{equation}
    \varepsilon_{urca}=N_A \lambda_{ec}E_{ec}+N_A \lambda_{\beta}E_{\beta}
    \label{e:e_urca}
\end{equation} 
where $\lambda_{ec}$ and $\lambda_{\beta}$ are the \ecap~ and the \betad~ rate, respectively, and $N_A$ is the Avogadro number. 

During the accretion phase, the activation of these weak interactions has two major consequences. First of all, because of the electron captures, the average neutronization of the WD increases. On the other hand, at the URCA shell there is a net energy loss due to the emission of $\nu\overline{\nu}$ pairs. 
When an Urca shell firstly forms at the center, the neutrino/antineutrino energy loss causes a sharp decrease of the central temperature (see, e.g., Figure 2 in MR2016, Figure 1 in P2017 and Figure 1 in S2017, where the effects of the activations of the \ecap~ on $^{23}$Na and $^{25}$Mg are shown). Then, as matter continues to be accreted, the mass coordinate of the URCA shell, as defined by the condition $\rho=\rho_{th}$, moves outward. This progressive displacement  of the URCA shell produces a uniform cooling of the core. Below the URCA shell, the abundance of the daughter nuclei increases, because the \ecap, until all the mother nuclei have been consumed. As a result, the matter inside a given URCA shell becomes $n$-rich. Once the 
maximum abundance of the $n$-rich isobar is attained, the \ecap s cease and the core heating, as due to the WD compression, restarts. 

As discussed in the extant literature, 
the most important isotopes involved in URCA processes during the accretion phase of \sna\ progenitors are \isotope{23}{(Na,Ne)}, \isotope{25}{(Mg,Na)} and the 
triplet \isotope{56}{(Fe,Mn,Cr)}\footnote{In this case, two consecutive \ecap s~ convert the mother \isotope{56}{Fe} into the daughter \isotope{56}{Cr}, through \isotope{56}{Mn}.}. 
Indeed, the abundances of the mother nuclei are, in these cases, large enough to produce sizable effects on the WD thermal content (see, 
\textit{e.g.}  P2017).
\begin{figure}
\plotone{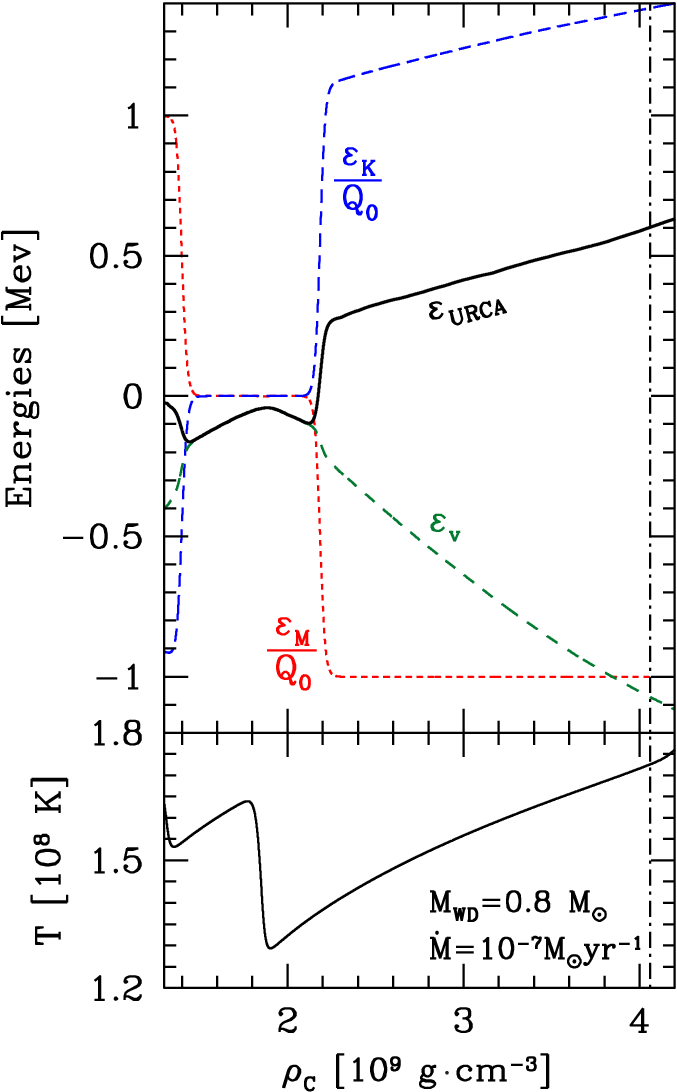}
\caption{Upper panel: Energy contributions from the \isotope{23}{(Na,Ne)} URCA pair at the center of a CO WD with initial mass 0.8 \msun\, accreting mass at 
\mdot=$10^{-7}$\msun\pyr. The term related to the mass excess $\varepsilon_{M}$ and to the kinetic energy of the eletrons $\varepsilon_{K}$ are normalized 
to the mass excess energy of the process \isotope{23}{Na}($e^-,\nu$)\isotope{23}{Ne} $Q_0$=4.3758 MeV. 
Lower panel: evolution in the $\rho-T$ plane of the center of the accreting WD.The dot-dashed vertical line marks C-ignition epoch, 
defined as the epoch when nuclear energy exactly balances energy losses via neutrino emission.}\label{f:e_urca}
\end{figure}

Later on, as the accreting WD approaches the Chandrasekhar mass, it undergoes a rapid compression, until physical conditions for C-burning are attained. 
The C-ignition occurs (near) the center and proceeds mainly through the $p$-channel \isotope{12}{C}(\isotope{12}{C},p)\isotope{23}{Na} and the $\alpha$-channel 
\isotope{12}{C}(\isotope{12}{C},$\alpha$)\isotope{20}{Ne}, the relative contribution being 0.44-0.45 and 0.56-0.55, respectively \citep[see, for instance, ][]{caughlan1988}.
As the $p$-channel starts to release \isotope{23}{Na}, it immediately undergoes an \ecap, thus producing a further increase of the local neutronization. The contribution of these \ecap s\, to the total nuclear energy budget can be estimated according to Eq.~\ref{e:eec}. Note that since the Fermi energy of the electrons is quite large, the \isotope{23}{Ne} daughters are produced in excited states and the energy released after the radiative decay exceeds that of the emitted neutrino. As a result, this process now produces a local heating. 
This is clearly shown in Figure~\ref{f:e_urca}, where we report for the \isotope{23}{(Na,Ne)} URCA shell the three contributions 
in Eq.~\ref{e:e_urca}, defined as
\begin{equation}
\varepsilon^i={\frac{n({^{23}Ne})\lambda_\beta E^i_{\beta}+n({^{23}Na})\lambda_{ec} E^i_{ec}}{n({^{23}Ne})\lambda_\beta+n({^{23}Na})\lambda_{ec}}}, 
\end{equation}
where the subscripts $\beta$ and $ec$ refer to the \betad\, and \ecap\, processes, respectively. As a matter of fact, the inclusion of URCA processes determines a larger energy 
release associated with C-burning at the center, so that the physical conditions for the thermonuclear runaway are attained sooner. 

Owing to the high degeneracy of the electron component, C-burning rapidly turns into a themonuclear runaway, so that the innermost zone of the accreting WD becomes 
unstable for convection and the \textit{simmering phase} begins. Moreover, due to the large release of energy at an increasing rate, the convective zone extends outwards, engulfing 
larger and larger portions of the WD. 
The role played by the \isotope{23}{(Na,Ne)} URCA shell during the simmering phase is not quite different from that played during the previos accretion phase until the external border of the convective-unstable core remains confined in the region where the density is larger than $\rho_{th}$. 
Then, convection smears off the ashes of the C-burning through the whole convective core, moving outward material enriched in $n$-rich isobars. When the border of the convective core approaches the 
\isotope{23}{(Na,Ne)} URCA shell, \isotope{23}{Ne} decay into the $n$-poor isobar \isotope{23}{Na}. As a result, \isotope{23}{Ne} is dredged up, meanwhile \isotope{23}{Na} is dredged down. 
If the convective mixing is fast enough, compared to the weak interaction timescales, this \isotope{23}{Na}-\isotope{23}{Ne} circulation implies a further cooling at the URCA shell, because of the antineutrino emitted by the \isotope{23}{Ne} decays, and a heating in the innermost zone, as due to the \ecap s~ on \isotope{23}{Na} nuclei.
  
Moreover, as firstly pointed out by \citet{iben1978b}, the central n-rich zone is also electron-poor, so that convection also moves downward the electrons released by the \betad s~ taking place at the URCA shell. 
In their downward motion, the thermal energy of the electrons should increase up to the local value of the Fermi energy. Therefore, the electron excess causes a cooling of the convective core. As already noted in 
\citet{iben1978b}, the electron abundance profile within the convective core critically depends on the mixing timescale, as compared to the local \ecap~ timescale. In order to take into account this energy contribution, Eq.~\ref{e:e_urca} should be modified as:
\begin{equation}
\varepsilon_{URCA}=N_A \lambda_{ec}E_{ec}+N_A \lambda_{\beta}E_{\beta}+\varepsilon_{W},
\label{e:e_urca2}
\end{equation}
where $\varepsilon_{W}$ is the negative contribution representing the amount of thermal energy absorbed when electrons are dredged down (see \S \ref{s:ibenw}).

Last, but not least important is the criterion adopted to define if a layer is stable or unstable against convection. In general, convection arises when the temperature gradient drops below a critical value. However, the molecular weight gradient may prevent convection (if negative) or may enhance the instability (if positive). 
According to the Ledoux criterion, a layer is stable against convection if:
\begin{equation}\label{e:ledoux_cr}
\nabla_{rad}<\nabla_{ad}+{\frac{\phi}{\delta}}\nabla_\mu, 
\end{equation}
where $\nabla_{rad}=\left({\frac{d\ln T}{d\ln P}}\right)$ is the radiative gradient,
$\nabla_{ad}=\left({\frac{\partial\ln T}{\partial\ln P}}\right)_S$ is the adiabatic gradient,
$\nabla_{\mu}=\left(\frac{d\ln \mu}{d\ln P}\right)$ is the molecular weight gradient,
$\phi=\left({\frac{\partial\ln \rho}{\partial\ln \mu}}\right)_{(P,T)}$ and 
$\delta=-\left({\frac{\partial\ln \rho}{\partial \ln T}}\right)_{(P,\mu)}$.
In case of a very efficient mixing, $\nabla_{\mu}$ vanishes, so that the Ledoux criterion reduces to the Schwarzschild criterion, i.e., $\nabla_{rad}<\nabla_{ad}$.
During the simmering phase, this condition is satisfied when the border of the convective core remains well inside the \isotope{23}{(Na,Ne)} URCA shell. On the contrary, when the convective-core border reaches the URCA shell, the rapid decay of the $n$-rich isobars conserve and reinforce the $\mu$-gradient, which represent a barrier to a further outward mixing. In practice, the growth in mass of the convective core stops when the convective border attains the location of the URCA shell \citep[see][for a detailed and complete analysis of this process]{lesaffre2006}.

\section{Critical analysis of the extant literature}\label{s:literature}

The analysis performed in the previous section gives us the possibility to critically review models for the simmering phase presented up to now (see \S~\ref{s:intro}), 
explaining the origin of the differences in the estimated neutronization at the explosion and evaluating their reliability. \citet{piro2008} provided the first evaluation of 
the neutronization evolution during the simmering phase. They started their computation from the onset of the simmering phase, ignoring the effects of URCA pairs/triplets 
in determining the physical conditions at the onset of central convection. However, they tried to account for such an effect  by varying the ($\rho$,T) starting values of their 
simulation. Moreover, even if they consider the WD structure, they practically used a one-zone model, evaluating nuclear burning at the center of the star and using the 
Schwartzschild criterion to define the temperature gradient (assumed to be adiabatic). In this way they were able to estimate the amount of \isotope{12}{C} consumed during the 
evolution up to the explosion and, hence, the produced neutronization. In this regard, they assume that convective mixing is instantaneous. The adopted nuclear network is 
very short, including 9 isotopes linked by 7 reactions only (see their Table 1); in particular they include one \pcap\, (\isotope{12}{C}($p,\gamma$)\isotope{13}{N}), 
one \acap\, (\isotope{13}{C}($\alpha,n$)\isotope{16}{O}) and one \ncap\, (\isotope{12}{C}($n,\gamma$)\isotope{13}{C}) only, which are relevant to evaluate the 
neutronization produced during the simmering phase but underestimate their contribution to the total energy budget. At the end, they assume that \ecap s onto \isotope{13}{N} and \isotope{23}{Na} 
do not contribute to the energy budget. In particular, this assumption is equivalent to set $\varepsilon_{URCA}=0$, because at the center of the CO WD during the simmering phase 
density is definitively larger than the threshold value for the \isotope{23}{(Na,Ne)} URCA shell so that the contribution from the \betad\, of \isotope{23}{Ne} could be neglected.
As a matter of fact, \citet{piro2008} ignore any energy contribution from convective URCA processes, only suggesting that if these processes are at work they should produce a cooling 
of the convective core so that a larger C-consumption is expected (see their footnote 1). 

On a general ground, the assumption that URCA processes do not contribute to the total energy budget during the simmering phase could be realistic, even if it can not be proved 
\textit{a priori}. Notwithstanding, the circulation via convective mixing of \isotope{23}{Na} and \isotope{23}{Ne} could produce local cooling and/or heating in the innermost zones 
where burning occur, so that the C-consumption estimated by \citet{piro2008} could be completely unrealistic and, consequentely, the neutronization at the explosion could be wrong. 

\citet{chamulak2008} used an approach similar to that of \citet{piro2008}, even if they used a full nuclear network including 430 isotopes and they computed the energy released 
by nuclear processes by including the mass term $\varepsilon_{M}$, the neutrino term $\varepsilon_{\nu}$ and the electron kinetic energy variation $\varepsilon_{K}$ (see their 
Eq.~2). As a consequence, they found that the \nane\, URCA process heats up the burning zones, because, as illustrated in Figure~\ref{f:e_urca}, for density values 
above $\sim2.1\times 10^9\mathrm{g\cdot cm^{-3}}$ the kinetic term dominates over the sum of the mass and neutrino terms ($\varepsilon_{URCA}\sim$ 1 MeV). 
The neutronization value they estimated at the explosion epoch is roughly similar to that by \citet{piro2008}, because both the two working group adopts similar assumptions 
and, in addition, the positive $\varepsilon_{URCA}$ from the URCA pairs \isotope{23}{(Na,Ne)} considered by \citet{chamulak2008} represent a $\sim$6\% correction to the 
effective energy delivered by the combustion of 6 \isotope{12}{C} isotopes.  

According to these considerations, the results by \citet{chamulak2008} suffer the same limitation as the previous ones: the assumption that URCA pairs provide an heating of the convective 
core can not be proved and, in addition, the role of convective URCA processes is ignored, as they can not be included in one-zone models. In this regard, the authors observed 
that the large abundance of the \isotope{23}{(Na,Ne)} URCA pairs could have a ``\textit{dramatic effect on the properties of the convection zone}''

MR2016 computed full evolutionary models of CO WD accreting matter, by considering different accretion rates, different initial thermal contents of the accreting WD
(\textit{i.e.} different initial central temperatures) and different initial metallicity of the Main Sequence WD progenitors. They adopted a nuclear network including 48 isotopes 
(up to sulfur) constructed by taking into account the results on the detailed nucleosynthesis computed in \citet{chamulak2008}. In their computation 
the equations describing the physical structure and the evolution of chemical species are solved simultaneously, accounting for the effects of convective mixing by means of a diffusion 
equation. The energy contribution related to URCA processes are computed according to Eq.~\ref{e:e_urca2}, by setting $\varepsilon_{W}$=0.
As they use a full evolutionary model, MR2016 can resolve the whole temperature and density profile during the accretion and the simmering phase, so that they can 
evaluate the contribution of URCA processes to the local and total energy budget of accreting WDs. Moreover, as convective mixing is directly included in the computation of models, 
they can also evaluate the effects of convective URCA processes on the neutronization at the explosion epoch. They show that, during the accretion the center of the star and, then, the 
innermost zones cools down due to the activation of URCA shell involving \isotope{25}{Mg} and \isotope{23}{Na}. 
They found, also, that the weak processes involving \isotope{23}{Na}
heat up the innermost zones of the accreting WD from the epoch of C-ignition and up to when the C-burning driven convection attains the \nane URCA shell. Moreover, 
as also detailed in S2017, they found that the mixing of processed matter through this URCA shell becomes an efficient mechanism to refurnish the innermost burning 
zone with fresh \isotope{23}{Na} and that the consequent \ecap s\, on this isotope represent a source of energy as important as carbon fusion reactions. The net result is that the 
amount of carbon consumed to trigger the explosion is lower than what estimated in one-zone models above and, hence, the neutronization at the dynamical break-out results lower. 

As already remarked, MR2016 assumed that the energy subtracted from the surround to carry down $e$-rich material from the URCA shell is negligible, an 
assumption that seems unrealistic, at least according to S2017, who computed \textit{a posteriori} the total energy associated to such a process and demonstrated that 
it is as large as the energy losses via neutrino emission in weak processes. Moreover, MR2016 did not mention what criterion they used to establish if a layer is 
convective unstable, even if it can be safely assumed that they use the Scwartzschild criterion \citep{rodri2019}. As a consequence they did not include the effects of possible 
molecular weight gradient on the properties of the convective zone. 

P2017 performed an analysis similar to that of MR2016, even if they used a different evolutionary code and a different nuclear network. They considered 
only one WD mass and a single value of the accretion rate and performed computations for three different initial metallicity of the Main Sequence progenitor of the WDs. As 
in MR2016 the physical structure equations and those describing the evolution of chemical species are solved simultaneously, by treating the convective mixing as an
advective process\footnote{For more detail about this issue see \S \ref{s:preanalysis}.}. In computing the evolution of accreting WDs they set $\varepsilon_{K}=\varepsilon_{W}=0$.
The adopted assumptions determine the same temperature and density profiles as in MR2016 during the accretion phase. In fact, as demonstrated above, for pre-existing URCA isobars, 
at the URCA shell the mass and kinetic energy contributions from mother and daughther nuclei have the same absolute value and cancel each other, the neutrino/antrineutrino sink 
remaining the only energy contribution. On the other hand, 
when carbon starts to be burnt in the innermost zones, P2017 estimated a negative energy contribution from the \nane\, URCA process in the burning region, while it should be 
positive (see Figure~\ref{f:e_urca}) so that they overestimated the amount of \isotope{12}{C} comsumed and, hence, the produced neutronization. At the end, during the simmering 
phase, their assumption about the energy contribution from the \nane\, URCA process is equivalent to assume that the energy subtracted to the surround to carry downward 
$e$-rich material is equal to the electrons kinetic energy released in \ecap s. Such an assumption could be valid on average, inside the convective core, but it is not locally valid. 
Moreover, such an assumption assume that the net results of URCA processes is a cooling of the accreting WD, whose effects is mimicked by arbitrarely reducing the amount 
of energy delivered in the burning region.
As a result they find definitively larger neutronization values at the explosion epoch and, in addition, that the amount of \isotope{12}{C} consumed during the simmering depends on 
the initial metallicity of the WDs progenitors. These results are the direct consequence of the assumption concerning the energy contribution of URCA processes. In fact, the 
larger the initial abundance of URCA isobars, the larger the cooling effects during the accretion phase and, therefore, the larger the density value at which C-burning occurs. 
Hence, due to the fact that 
$\varepsilon_{\nu}$ does depend mainly on density, the larger the initial Z, the more efficient the neutrino cooling and, hence, the larger C-consumption and final neutronization. 

\section{Preliminary analysis}\label{s:preanalysis}

According to previous considerations, it comes out that all the models in the extant literature for the simmering phase of SNe Ia are not accurate in reproducing the physical and 
chemical processes at work in real accreting WDs. In fact, all of them are based on oversimplified assumptions that can not be proved \textit{a priori} and that, in some 
cases, are incorrect. The previous analysis has remarked once again the evidence that the neutronization level at the explosion epoch depends mainly on the energy contribution of 
URCA isotopes during the simmering phase which depends not only on local properties but also on the ouward and inward circulation of $n$-rich and $e$-rich material, 
respectively, the efficiency of the latter process depending critically on the possibility that the convective zone could encompass the \nane\, URCA shell. 

In the following we presents models including all the energy term in Eq.~\ref{e:e_urca}, by paying particular attention to the role played by the formation of a strong 
$\mu$-gradient at the URCA shell and to its effects on the physical and chemical properties of the zone unstable for convection. In our simulation we make use of the FuNS code 
as in P2017 \citep{straniero2006,cris2009}. The nuclear network we adopt is slightly different from that in P2017 and it includes 200 nuclear processes linking 50 isotopes, 
namely $p,\ \alpha,\ n$, C(12-14), N(13-15), O(16-20), F(18,22), Ne(20-24), Na(22-25), Mg(23-28), Al(25-28), Si(27-32), P(30-32), \isotope{56}{Mn}, \isotope{56}{Cr}, \isotope{56}{Fe}.
With respect to P2017, we neglect isotopes above phosporus because, according to the extant literature \citep[\textit{e.g.}, see ][]{chamulak2008,piersanti2017} they are 
scarcely produced during the simmering phase. In doing so, we also neglect the contribution by the \isotope{32}{(Si,S,P)}, \isotope{37}{(S,Cl)} and \isotope{39}{(Ar,K)} 
URCA processes, because either their abundances are very low or they occur at densities so low that they provide a negligible contribution to both the total neutronization and the 
energetic of the accreting WD \citep[see Table 1 in ][]{piersanti2017}. The only exception is the URCA triplet \isotope{32}{(Si,S,P)} whose contribution to the neutronization along 
the whole accreting WD is $\delta\eta_c\sim 2\times 10^{-4}$. On the other hand, we include the \isotope{28}{(Si,Al,Mg)} and the \isotope{24}{(Mg,Na,Ne)} URCA triplets which 
plays an important role in determining both the ignition conditions and the neutronization of the accreting WDs. Reaction rates for these two URCA triplets are derived from \cite{suzuki2016},
while those for all the other URCA processes considered in the nuclear network are included as detailed in Paper I (see Section 2 and Table 1). Reaction rates for charged particles interactions 
are derived from the JINA REACLIB database \citep{cyburt2010}.

\begin{figure}
\plotone{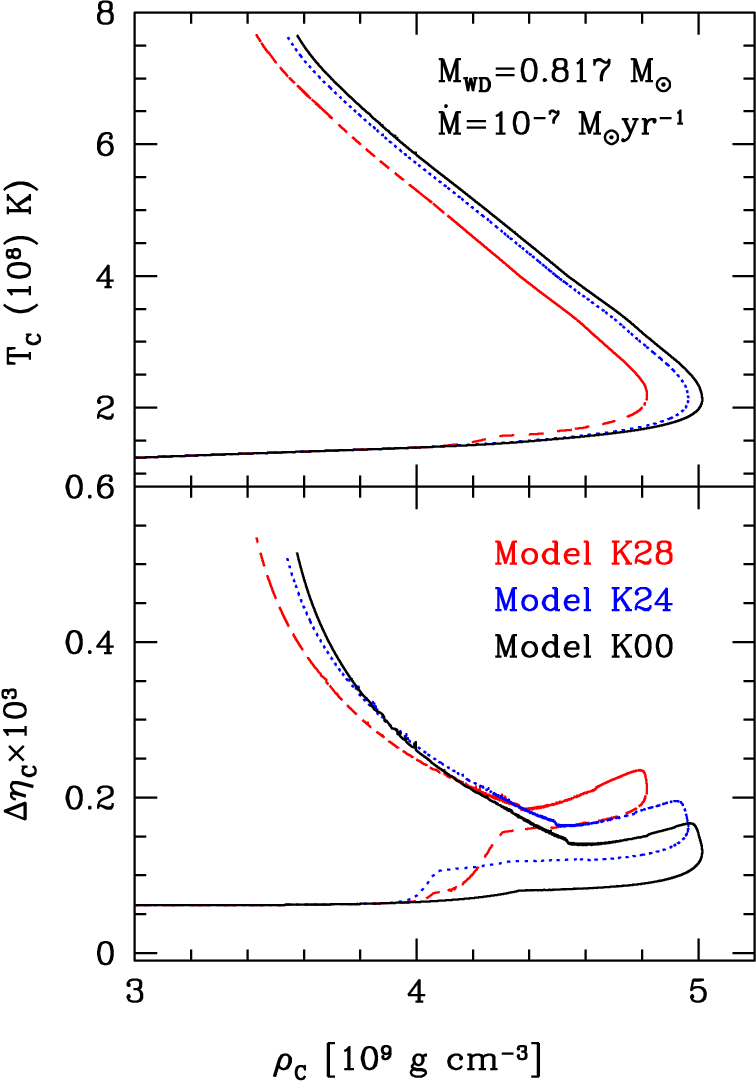}
\caption{Upper panel: Evolution in the $\rho-T$ plane of the center of the ZSUN CO WD accreting matter at \mdot=$10^{-7}$\msun\pyr. With respect to Model K00 (solid black 
line), Model K28 (dashed red line) and Model K24 (blue dotted line) include the \isotope{28}{(Si,Al,Mg)} and the \isotope{24}{(Mg,Na,Ne)} URCA triplets, respectively. 
Lower panel: Evolution of the neutronitazion variation $\Delta\eta_C$ at the center for the same models as in upper panel.}\label{f:new_urca}
\end{figure}

To illustrate how the \isotope{28}{(Si,Al,Mg)} and the \isotope{24}{(Mg,Na,Ne)} URCA triplets affect the evolution during the accretion and the simmering phases, 
we compute a set of test models by adopting the same network as in \citet{piro2008}, but including also the \isotope{25}{(Mg,Na)} and \isotope{56}{(Fe,Mn,Cr)} 
URCA processes. The energy contribution related to URCA processes is implemented according to Eq.~\ref{e:e_urca} during both the accretion and the simmering phase and the convective 
mixing is modelled as a diffusive process, so that the equation describing the physical structure and the evolution of chemical species can be solved simultaneously. 
The initial WD structure is the same ZSUN model in P2017 (\textit{i.e.} a CO WD 
with M=0.817 \msun, whose progenitor star had solar metallicity $\mathrm{Z_\odot=1.38\times 10^{-2}}$ - see P2017). CO-rich matter is accreted directly at 
\mdot=$10^{-7}$\msun\pyr. The results relative to this reference model (Model K00 - solid black line) are displayed 
in Figure~\ref{f:new_urca}, where we plot the evolution of the center of the accreting WD in the $\rho-T$ plane (upper panel) and the increase of neutronization at the center 
$\Delta\eta_C$ as a function of the central density (lower panel). This model is similar to those presented in MR2016 ($\mathrm{Z=Z_\odot}$ and \mdot=$10^{-7}$\msun\pyr) and 
in S2017, and has a quite similar evolution (for instance compare the lower panel in Figure~\ref{f:new_urca} and Figure 2 in S2017), even if the increase of the 
central neutronization at the epoch of dynamical breakout is larger (see also Table~\ref{t:newurca}). This is due partly to the inclusion in Model K00 of the \isotope{56}{(Fe,Mn,Cr)} 
URCA triplets which produce a variation of the neutronization of $\delta\eta_C= 2 {\rm X({{^{56}}Fe})/A({{^{56}}Fe})}\simeq 3.75\times 10^{-5}$, where we use 
X(\isotope{56}{Fe})=1.05$\times 10^{-3}$ (see Table 1 in P2017). The remaining discrepancy ($\delta\eta_{disc}\simeq1.56\times 10^{-4}$) has to be ascribed to the different nuclear 
network adopted, in particular to the fact that the very short network used here largely underestimates the energy contribution from $p$- and \acap\, and, hence a larger amount of 
\isotope{12}{C} has to be burnt in order to attain the explosion. 

In computing model K28 we include in the nuclear network also the \isotope{28}{(Si,Al,Mg)} URCA triplets. We recall that \ecap\, on \isotope{28}{Al} have a threshold density of 
$\rho_{th}\simeq 0.89\times 10^9$\gcc, while for \isotope{28}{Si} it is $\rho_{th}\simeq 4.24\times 10^9$\gcc. This means that when the physical conditions suitable for 
\ecap\, on \isotope{28}{Si} are attained the daughter nuclei \isotope{28}{Al} immediately capture an electron, forming \isotope{28}{Mg}. As a matter of fact, we find that, 
when the central density of the accreting WD exceeds $4.17\times 10^9$\gcc, already 10\% of the initial local \isotope{28}{Si} have been converted into \isotope{28}{Mg}. 
At that epoch the chemical potential of the electrons, or the kinetic energy of electrons, involved in \ecap\, is equal to $\varepsilon_K\sim$6.011 MeV while the mass excess 
energy term for the \isotope{28}{Al}($\mathrm{e^-,\nu}$)\isotope{28}{Mg} and the energy of the emitted neutrino are $\varepsilon_{M}$=-1.832 MeV and 
$\varepsilon_{\nu}$=1.981 MeV, respectively. This means that the activation of \isotope{28}{(Si,Al,Mg)} URCA process determines a rapid heating of the innermost zone of the accreting WD, 
as it can be clearly seen in Figure~\ref{f:new_urca}. Due to the larger central temperature, C-burning proceed at an higher rate, so that the innermost zones of the accreting WDs 
become unstable for convection at lower density (see Table~\ref{t:newurca}). After the onset of convection, the \isotope{28}{(Si,Al,Mg)} URCA triplet do not contribute anymore to both 
the energy budget during the simmering phase, because the density threshold to activate the $\beta$-decay of \isotope{28}{Mg} is too low. The conversion of the initial 
\isotope{28}{Si} into \isotope{28}{Mg} produces an increase of the local neutronization at the center of 
$\delta\eta_c=2 {\rm X({{^{28}}Si})/A(X{{^{28}}Si})}\simeq 5.21\times 10^{-5}$. 
This explains the larger $\Delta\eta_C$, with respect to model K00, obtained at the epoch when convection attains the \isotope{23}{(Na,Ne)} URCA shell. However, since the 
accreting WD is globally less dense, a smaller amount of carbon has to be nuclearly processed in order to trigger the explosion. As a consequence, the value of central neutronization 
at the epoch of dynamical breakout is larger by only $\sim$4\% than that in the K00 model. 
\begin{deluxetable}{lrrr}
\centering
\tablecaption{Physical and chemical properties of models computed by accreting CO-rich matter at \mdot=$10^{-7}$\msun\pyr, but using different nuclear network. \label{t:newurca}}
\tablehead{
\colhead{Model} & \colhead{Model K00} & \colhead{Model K28} & \colhead{Model K24} 
}
\startdata
$\mathrm{\rho_{ign}}$ ($10^9$\gcc)         & 4.105 & 4.077 & 4.070 \\
$\mathrm{T_{ign}}$ ($10^8$ K)              & 1.732 & 1.728 & 1.727 \\
$\mathrm{\rho_{simm}}$ ($10^9$\gcc)        & 4.495 & 4.187 & 4.459 \\
$\mathrm{T_{simm}}$ ($10^8$ K)             & 1.819 & 1.751 & 1.818 \\
$\Delta$M(\isotope{12}{C})($10^{-2}$\msun) & 1.324 & 1.313 & 1.233 \\
$\mathrm{M^{max}_{conv}}$ (\msun)          & 1.213 & 1.213 & 1.213 \\
$\mathrm{\rho_{exp}}$ ($10^9$\gcc)         & 3.704 & 3.560 & 3.667 \\
$\mathrm{\Delta\eta_{c} (10^{-3})}$            & 0.544 & 0.563 & 0.540 \\
\hline                                                                              
\enddata
\tablecomments{${\rm\rho_{ign}}$ and ${\rm T_{ign}}$: central density  
and temperature at C-ignition; ${\rm\rho_{simm}}$ and ${\rm T_{simm}}$: 
central density and temperature at the onset of convection; 
$\Delta$M(\isotope{12}{C}): amount of 
\isotope{12}{C} consumed via nuclear burning from the onset of mass accretion up to the explosion;
${\rm M^{max}_{conv}}$: maximum extension of the convective zone; 
$\Delta\eta_{\rm c}$: variation of the neutronization at the center from the beginning of the accretion process
up to the explosion.}
\end{deluxetable}

The situation is quite different for the K24 model, which includes the \isotope{24}{(Mg,Na,Ne)} URCA triplet.  In this case the density threshold for \ecap\, on \isotope{24}{Na} ($\rho_{th}=5.07\times 10^9$\gcc) is 
slightly larger than that for the \ecap\, on \isotope{24}{Mg} ($\rho_{th}\simeq 4.16\times 10^9$\gcc. As a consequence, the activation of the \isotope{24}{(Mg,Na)} processes does not 
produce any sizable effects on the physical conditions at the onset of convection, but the whole \isotope{24}{(Mg,Na,Ne)} URCA triplet behaves as a heating process of the 
innermost zone of the accreting WD at the beginning of the simmering phase, due to the circulation of this isotopes forced by convective mixing. As a whole model K24 results less dense 
for a fixed temperature value, while the contribution of the considered URCA triplets to the central neutronization is 
$\delta\eta_c=2 {\rm X({{^{24}}Mg})/A({{^{24}}Mg})}\simeq 0.39\times 10^{-5}$. Also in this case, as discussed above for the \isotope{28}{(Si,Al,Mg)} URCA triplet, a smaller 
amount of \isotope{12}{C} has to be burnt to produce the explosion as the structure is globally less dense. As a consequence, the central neutronization at the breakout epoch is practically the same 
as in model K00 (only $\sim0.7$\% lower). 

The other important difference with respect to P2017 is the treatment of convective mixing as a diffusive process, whereas in that paper it was modelled according to an advective scheme. 
S2017 partially analyze the effects of different scheme on the neutronization level at the epoch of the explosion. To this aim, they assume that the ``advective scheme'' in 
P2017 corresponds to the mixing algorithm used in the FuNS code as described in \citet{straniero2006}. Such an algorithm was originaly introduced by \citet{sparks1980}, 
successively corrected by \citet{chieffi1998} and linearized in the form currently avalilable in the FuNS code by \citet{straniero2006}. Hereinafter we refer to this mixing algorithm 
as SEL (\textit{Sparks-Endal-Linearized}). Such an algorithm evaluates the local variation of chemical composition via convective mixing as due to effective couple of a 
given layer with all the others in the convective-unstable zone. As a consequence the SEL mixing scheme can not be implemented and solved together with the equations describing 
the physical structure and nuclear burning along the star and, for this reason, the operator splitting technique has to be employed. S2017 found that the central 
neutronization at the epoch of dynamical breakout is a factor $\sim2$ larger in the model based on the SEL scheme and explained such a discrepancy as a consequence of 
the different mixing efficiency before the occurrence of the mixing freeze-out \citep{nonaka2012}.

\begin{deluxetable*}{lrrrrr}
\tablecaption{Model inputs and results. \label{t:mixsch}}
\tablehead{
\colhead{Model} & \colhead{Model K00} & \colhead{Model SEL} & \colhead{Model DFd} & \colhead{Model ADd} & \colhead{Model ADV} 
}
\startdata
$\mathrm{M^{max}_{conv}}$ (\msun)          & 1.213 & 1.241 & 1.211 & 1.209 & 1.258 \\
$\mathrm{\Delta\eta_{c} (10^{-3})}$            & 0.544 & 1.702 & 0.584 & 2.081 & 0.211 \\
$\mathrm{\overline{\eta}_{exp} (10^{-3})}$  & 0.322 & 0.403 & 0.355 & 0.417 & 0.287 \\
$\mathrm{\rho_{exp}}$ ($10^9$\gcc)            & 3.704 & 3.649 & 3.729 & 3.810 & 1.942 \\
$\Delta$M(\isotope{12}{C})($10^{-2}$\msun) & 1.324 & 1.550 & 1.478 & 1.596 & 1.112 \\
\hline                                                                              
\enddata
\tablecomments{
${\rm M^{max}_{conv}}$: maximum extension of the convective zone; 
$\Delta\eta_{\rm c}$: variation of the neutronization at the center from the beginning of the accretion processup to the explosion; 
${\overline{\eta}}_{\rm exp}$: neutronization at the explosion averaged over the convective zone ${\rm M^{max}_{conv}}$;
$\mathrm{\rho_{exp}}$: density at the explosion epoch;
$\Delta$M(\isotope{12}{C}): amount of \isotope{12}{C} consumed via nuclear burning from the onset of mass accretion up to the explosion.}
\end{deluxetable*}

We remark that the mixing algorithm used in P2017 is not the SEL scheme assumed by S2017, but is is based on the 
assumption that the transport of matter by both the forward and the downward convective fluxes are regulated by an 
advective equation, namely:
\begin{equation}
\frac{\partial X}{\partial t}+4\pi r^2\rho v\frac{\partial X}{\partial m}=0
\label{e:advsch}
\end{equation}
where $m$ and $t$ are the mass coordinate and time, while $\rho(m,t)$, $v(m,t)$ and $X_j(m,t)$ are density, average 
convective velocity and  mass fraction of the isotope $j$,  respectively.
As in case of diffusion, also the advective scheme can be arranged in order to solve simultaneously all the relevant 
equations describing the physical and chemical properties of the accreting WD. Nevertheless, advection and diffusion 
describe different physical processes. In fact, diffusion often refers to microscopic transport of particles 
(or heat), while advection describes a bulk transport of mass (or heat). In general, convection combines diffusion 
and advection, but the vertical flow, as driven by the thermal gradient, is mainly due to advection.

Currently, a self-consistent description of turbulent convection is still missing. The choice of the scheme used to 
describe this process is a critical issue of the extant models of \sne~ progenitors. Indeed,
different mixing algorithms produce different mixing efficiencies and, during the simmering phase, the
energy contribution from the URCA processes is taken into account according to Eq.~\ref{e:e_urca2}, whose 
value may be positive or negative, depending on the circulation efficiency of the URCA isobars.
\begin{figure}
\plotone{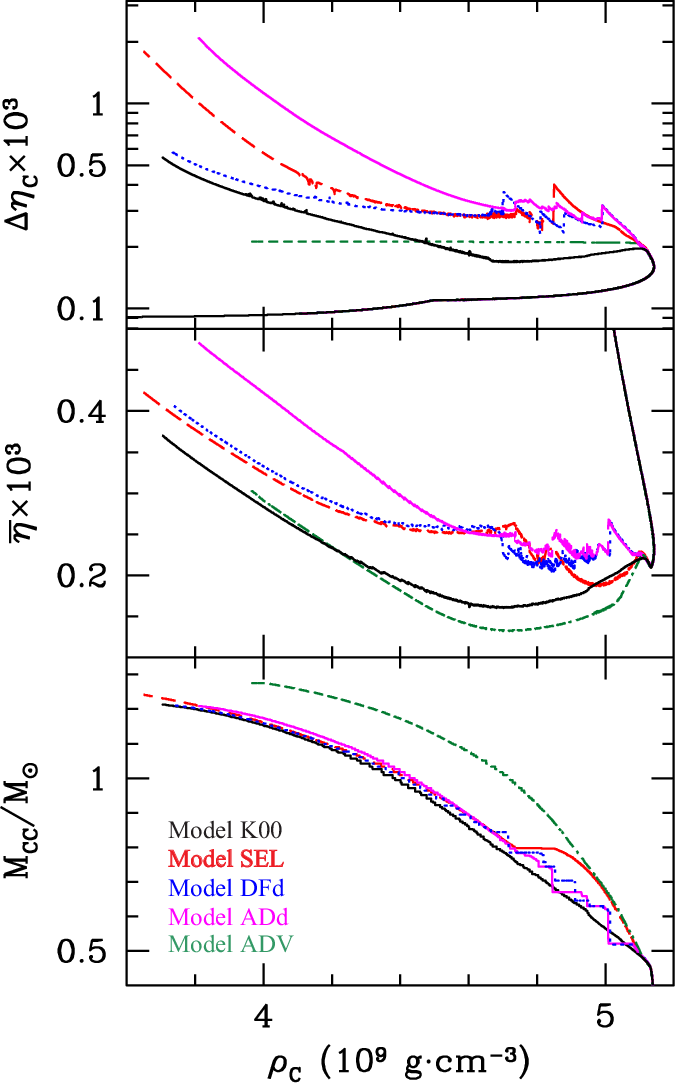}
\caption{Evolution as a function of the central density $\rho_C$ of the neutronization variation at the center (upper panel) of the average neutronization inside the 
convective-unstable region (middle panel) and of the maximum extention of the convective unstable region (lower panel). Different lines refer to models computed with 
different mixing algorithm and adopting or not the operator splitting technique to treat convective mixing. Solid black: Model K00; long-dashed red line: Model SEL; dotted blue 
line: Model DFd; dot-dashed magenta line : Model ADd; dashed green line: Model ADV.\label{f:mixsch}}
\end{figure}

To investigate more in detail such an issue, we compute a second set of test models, by assuming the same short network as in model K00 above, but varying the mixing algorithm. 
In particular, we compute models by using in the modellization of convective mixing the SEL, advective and diffusive schemes (models SEL, ADd and DFd, respectively), 
but in all the three cases we apply the operator splitting technique and compute the effect of mixing after each time-step integration. 
At the end, we compute also a model by using the advective scheme and solving simultaneously all the equation describing the evolving WD (model ADV). The results are 
displayed in Figure~\ref{f:mixsch} and some relevant quantities of the computed models are listed in Table~\ref{t:mixsch}.
Our computations confirm previous results from S2017 that the SEL mixing scheme predicts at the explosion epoch values for $\eta_C$ definitively larger with respect to model K00;  
at the same time we found that also model DFd has a slightly larger ($\sim$ 7\%) value of $\eta_C$. Moreover, Figure~\ref{f:mixsch} suggests that the evolution 
of all ``decoupled'' models (SEL, DFd and ADd) after convection has attained the \isotope{23}{(Na,Ne)} URCA shell is quite different from that of model K00. This clearly indicates 
that the operator splitting technique used to compute the effects of convective mixing underestimates the effects of heating related to the circulation of URCA isobars, 
because their energetic feedback is not consistently taken into account, and, hence, automatically predicts larger neutronization values. The differences existing in the final value of 
$\eta_C$ among these models with different mixing scheme has to be ascribed to different mixing efficiency, the diffusive approach having the highest one and the advective one 
the lowest, respectively. 

\begin{figure}
\plotone{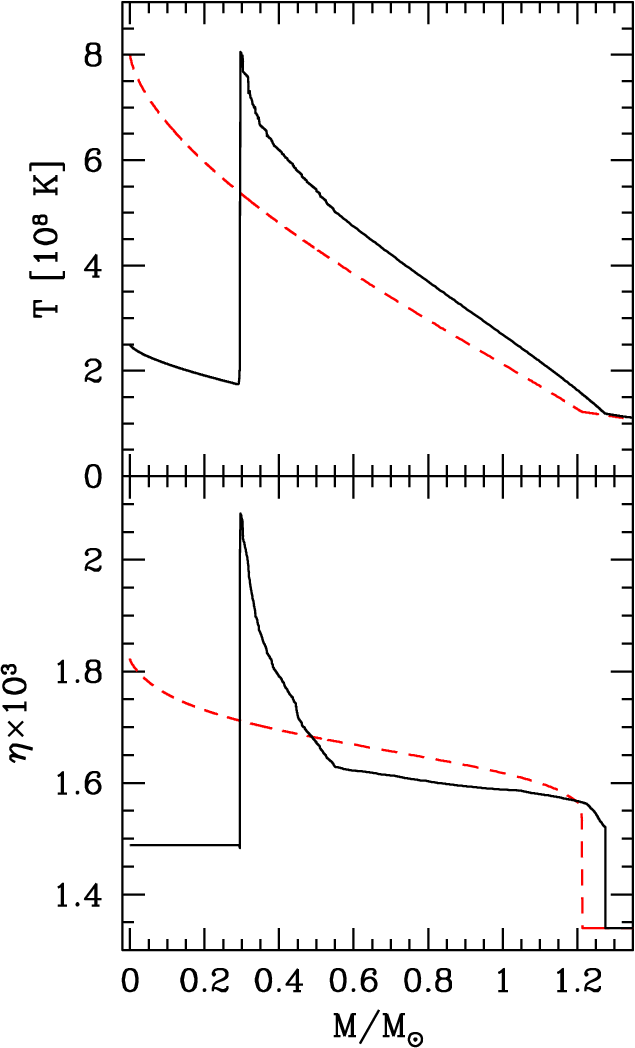}
\caption{Temperature (upper panel) and neutronization (lower panel) profile for model ADV at the explosion epoch. For comparison the same quantities are also plotted 
for model K00 (dashed red line).\label{f:advmod}}
\end{figure}

The evolution of model ADV, \textit{i.e.} with convective mixing modeled according to Eq.~\ref{e:advsch} which is solved contemporary to all the other equations describing the physical 
structure and nuclear burning, is completely different. In fact, after central convection has attained the \isotope{23}{(Na,Ne)} URCA shell, due to the low mixing efficiency, the 
dredged-down \isotope{23}{Na} undergoes \ecap\, during the inward motion well before it could attain the center of the accreting WD. This produce a local heating at the 
mass coordinate $M_{sp}\sim$ 0.295\msun\, and, very soon, the convective core splits into two convective region: a innermost one, extending from the center to $M_{sp}$ and 
outermost one up to the \isotope{23}{(Na,Ne)} URCA shell. In the outer convective shell the circulation of URCA isobars continues to produce an efficient heating, 
while the C-burning rate progressively increases close to $M_{sp}$ as the local temperature increases, so that the neutronization in this zone remains practically constant. As temperature 
increases at $M_{sp}$, the convective shell encompasses the \isotope{23}{(Na,Ne)} URCA shell and progressively engulfs larger and larger portion of the accreting WD.
On the other hand, in the convective core, the circulation 
of URCA isobars does not produce any energy contribution because the outer border of the convective core is well confined inside the corresponding URCA shell. As a 
consequence $\eta_C$ increases very slowly, as due to C-burning and subsequent production of \isotope{23}{Na}. When the timescale for \ecap\, on \isotope{23}{Na} becomes 
longer than the local heating timescale at $M_{sp}$, neutronization in the convective shell starts to increase due to \ecap\, on \isotope{13}{N}. Therefore, at the explosion 
epoch both the temperature and neutronization profile for models exhibit a peak at $M_{sp}$, as displayed in Figure~\ref{f:advmod}. Such a result is a consequence only of the 
mixing efficiency related to the advective mixing scheme. As a test we arbitrarely increase by a factor of 50 the convective velocity inside the convective-unstable zone and 
we find that the original convective core splits at the mass coordinate $M^{50}_{sp}\simeq$ 0.21\msun, thus confirming the effects of the adoped mixing scheme 
(\textit{i.e.} the adopted mixing efficiency) on the physical and chemical properties of the simmering phase.

\section{Next Generation Models}\label{s:ngmodels}

\subsection{Inward motion of e-rich material}\label{s:ibenw}
We continue our analysis of the simmering phase of \sne\, by using as case study the model ZSUN introduced in previous section accreting CO-rich matter at \mdot=$10^{-7}$\msun\pyr. 
We adopt the diffusive scheme for convective mixing and we use the full nuclear network (including 52 isotopes linked by 200 nuclear processes), except the weak 
processes associated with the URCA \isotope{21}{(F,Ne)} isobars.
We compute first a reference model (label REF) by including the energy contribution 
from URCA isotopes according to Eq.~\ref{e:e_urca}. Then we compute a second model by including the cooling effects due to the inward motion of $e$-rich material in zones 
unstable for convection according to Eq.~\ref{e:e_urca2}. As already recalled, the importance of the contribution $\varepsilon_W$ was first discussed by \citet{iben1978b} who suggested 
that it should be proportional to effective inward flux of electron and to the variation of the electronic chemical potential $\mu_e$, \textit{i.e.} the Fermi energy, along the 
inward motion. In formula \citep[see also Eq. 8 in ][]{iben1978b}:
\begin{equation}
\varepsilon_W=-{\frac{\partial\mu_e}{\partial M}}\Phi_e(M)
\end{equation}
where $\Phi_e(M)$ is the inward flux of electrons through a sphere of mass $M$ and it is defined as:
\begin{equation}
\Phi_e(M)=-N_A\int_0^M\left({\frac{dY_e}{dt}}-{\frac{d\bar{Y}_e}{dt}}\right)dm
\label{e:flux_iben}
\end{equation}
where $N_A$ is the Avogadro's number, $\frac{dY_e}{dt}$ the local variation of $Y_e$ as determined by weak processes and $\frac{d\bar{Y}_e}{dt}$ \textit{the mass-averaged 
(and therefore actual local) rate of change of the electron abundance in the core} as determined by both convection and nuclear burning. 

As we directly couple the solution of all the equations describing the physical and chemical structure of the accreting WD we can release the assumption that the chemical composition 
in the convective core is homogeneous, so that in implementing Eq.~\ref{e:flux_iben} we can include directly the local variation of the free electrons number:
\begin{eqnarray}\label{e:flux_our}
\Phi_e(M)&=&-N_A\int_0^M \left[\left({\frac{dY_e}{dt}}\right)_{nucl}-{\frac{\delta Y_e(m)}{\delta t}}\right]dm \\\nonumber 
{\frac{\delta Y_e(m)}{\delta t}}&=&{\frac{Y_e(t_0+\delta t,m)-Y_e(t_0,m)}{\delta t}}
\end{eqnarray}
where $\delta t$ is the actual timestep on which the new structure is integrated. 
S2017 correctly observed that, when the effects of convective mixing are included in the iteration procedure, the argument of the integral in Eq.~\ref{e:flux_iben}, reduces to 
the mixing term only, so that by using the diffusion scheme to model convective mixing, it becomes:
\begin{equation}
\Phi_e(M)=N_A\left(4\pi r^2(M)\rho(M)\right)^2D(M){\frac{\partial Y_e}{\partial m}}\biggr|_M
\label{e:flux_s2017}
\end{equation}
where $D(M)$ is the diffusion coefficient at $M$. 

We verified that both the evaluations for $\Phi_e(M)$ in Eq.~\ref{e:flux_our}, \ref{e:flux_s2017} produce the same results when the $\varepsilon_W$ energy term  is 
evaluated \textit{a posteriori} after the convergence of model at each timestep.
However, when we try to incorporate Eq.~\ref{e:flux_s2017} in the model computation we found that this formulation largely overestimated the inward electron flux, so that 
the corresponding energy contribution $\varepsilon_{W}$ becomes larger and larger, making the computation impossible. On the other hand, when we use Eq.~\ref{e:flux_our} we are 
able to perform the integration of models. The reason for such an occurrence is not related to inconsistent assumptions in deriving the above formulation but only to how 
Eq.~\ref{e:flux_s2017} has been implemented. In fact, in our code the border of convective-unstable zones and the mixing efficiency there (\textit{i.e.} the diffusion coefficients) are 
determined at the beginning of the iteration procedure and then are maintained constant. This procedure is commonly adopted in the computation of stellar models 
to avoid numerical fluctuations during the integration procedure \citep[see, for instance, Section 6.2 in][]{paxton2011}. In the computation of $\Phi_e(M)$, however, 
such a procedure introduce an inconsistency, as the mixing term at the beginning of a new iteration step is overestimated with respect to the previous converged solution. This 
occurrence is particularly relevant at the external border of the convective core where new zones radiative at a previous time become now unstable for convection. As a consequence, 
the local inward electron flux is very large (in the previous integrated model the mixing efficiency there was zero) and this drives the model very far from the convergence 
domain.
\begin{figure}
\plotone{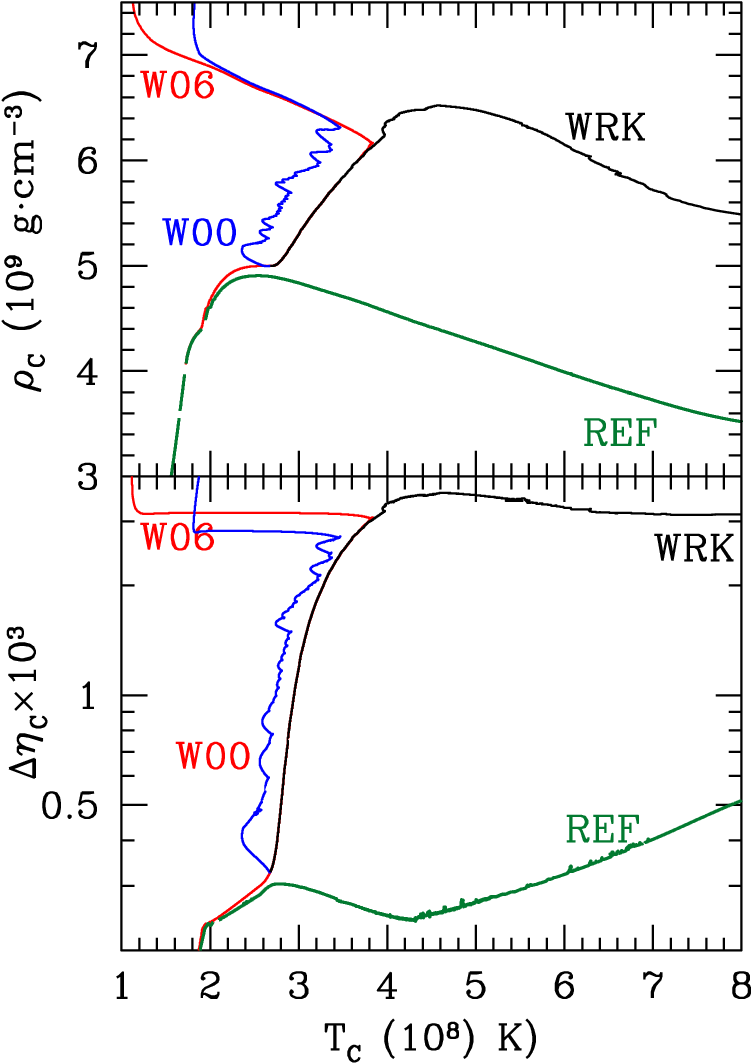}
\caption{Evolution of the density (upper panel) and of the neutronization variation (lower panel) as a function of the central temperature. All the displayed models include the 
energy contribution due to the mixing of $e$-rich material from the outer zones of the convective core downward (see Eq.~\ref{e:e_urca2}). Model W00 has been computed 
with a very coarse mass grid everywhere, while for model W06 we adopt a very high mass-resolution in the innermost 0.6\msun\, zone and a low mass-resolution above. 
For comparison model REF, implementing the energy contribution from URCA pairs as in Eq.~\ref{e:e_urca}, is also displayed.\label{f:ibenmod}}
\end{figure}

In his seminal work, \citet{iben1978b} found that when the border of the convective core approaches the \isotope{23}{(Na, Ne)} URCA shell, the accreting WD undergoes thermal 
oscillations. This is the consequence of the rapid increase of the neutrino/antineutrino emission from the \isotope{23}{(Na,Ne)} URCA pairs due to both the increased abundance of 
these isotopes (they are produced by C-burning) and the increased density at the center which determines a larger cross section for these weak processes. When the cooling effects 
related to URCA processes becomes the dominant energy losses in the convective core, thermal oscillations starts and proceed through a sequence of alternate cooling and heating phases.
In the former, the URCA neutrino/antrineutrino cooling is larger than the heating via nuclear burning and release of the kinetic energy of electrons from \ecap; as a consequence, 
the temperature in the whole convective core decreases, the mass of the convective unstable zone reduces and the structure undergoes a compression. In the latter phase, 
due to decrease of URCA neutrino/antineutrino cooling, heating becomes the dominant process, so that the temperature in the core rises again and the convective-unstable 
region grows in mass again. 

Our results are displayed in Figure~\ref{f:ibenmod}, where we report as a function of the central temperature the evolution of the density (upper panel) and neutronization 
(lower panel) at the center. The model implementing Eq.~\ref{e:e_urca2} is labeled WRK (solid black line); for comparison we also plot data for the model REF (long-dashed 
heavy green line). As it can be seen in the upper panel of Figure~\ref{f:ibenmod}, thermal oscillations do not occur, or, better, their amplitude is very small, even if the general trend 
reported by \citet{iben1978b} (slow increase of central temperature and rapid compression of the whole core) is confirmed . Such an occurrence is a direct 
consequence of the very high resolution in mass adopted in our computation (meshes in the adopted mass grid  have mass extension $\Delta M$ lower than $5\times 10^{-4}$\msun) 
and this is confirmed by comparing model WRK with the W00 model (blue line in Figure~\ref{f:ibenmod}), which was computed by adopting a very coarse mass grid 
($\Delta M\ge 10^{-2}$\msun) . As an additional test, we computed model W06 (red line in Figure~\ref{f:ibenmod}) by adopting a very fine mass grid in the innermost 0.6\msun\, zone 
($\Delta M\le 5\times 10^{-4}$\msun) of the accreting WD and a coarse one outside ($\Delta M\ge 10^{-2}$\msun). As it can be seen, when the mass of the convective core exceeds 
0.6\msun, a huge thermal oscillation occurs. Hence, we conclude that the characteristics of thermal oscillations depend on the adopted mass grid for the computation not 
only close to the \isotope{23}{(Na,Ne)} URCA shell but in the whole convective core. 

\citet{iben1982b} investigated also the behaviour of CO WD accreting H-rich matter; as in his previous work devoted to analyze the effects of URCA processes in the evolution 
of CO core undergoing C-burning, he stopped the computation well before the final outcome could be assessed. Notwithstanding, he suggested that thermonuclear runaway 
triggered by C-burning at the center can not be limited by the cooling effects at the URCA shell, so that the net results of including the cooling effects produced by inward convective mixing 
of $e$-rich matter is only a delay of the dynamic explosive event. Our simulation (model WRK in Figure~\ref{f:ibenmod}) confirms such a guess, as we obtain that, during the phase 
of slowly increasing temperature, the total luminosity above the \nane\, URCA shell progressively increases from negative values up to zero. In this condition, convective-unstable 
core can encompass the URCA shell and the heating at the center rapidly accelerates up to the explosion. However, we also found that during the ``thermal oscillation'' phase, 
due to the reduced efficiency of heating via circulation of $e$-rich material, the evolution is mainly driven by nuclear burning of \isotope{12}{C}; this implies a large production of 
\isotope{23}{Na}, and, hence, a large increase of neutronization due to the \isotope{23}{Na}(e$^{-},\nu$)\isotope{23}{Ne} weak process. This is clearly displayed 
in the lower panel of Figure~\ref{f:ibenmod},  where we report the variation of the neutronization at the center. As it can be seen, during the ``thermal oscillation'' phase $\eta_C$ 
rapidly increases, while later on, it remains almost flat and then it decreases due to the fact that convective core encompass the URCA shell and the timescale for 
\ecap\, on \isotope{23}{Na} becomes shorter that the local heating timescale. As a matter of fact, we find that at the dynamical breakout, model WRK has 
$\mathrm{\rho_{exp}=5.47\times 10^9}$\gcc, 
$\mathrm{M_{conv}^{max}=1.065}$\msun,
$\mathrm{\Delta M({^{12}C})=4.83\times 10^{-2}}$\msun,
$\mathrm{\Delta\eta_C=42.86\times 10^{-3}}$, a factor 8 larger with respect to model REF.

\subsection{Criterion for convection}\label{s:convcrit}
As it is well known, stellar matter undergoing nuclear reactions evolves progressively from light isotopes to heavy ones, determining a continuous increases of the molecular weight $\mu$.  
If nuclear burning occurs in a convective zone, as in the convective core of accreting WDs during the simmering phase, at the borders of this zone a gradient of molecular weight 
$\nabla_\mu=d\log\mu/d\log P$ naturally forms and the temperature gradient and the position of the borders of the layers unstable for convection should be evaluated not on 
the base of the Schwartzschild criterion but according to the Ledoux criterion in Eq.~\ref{e:ledoux_cr}. As the ratio $\phi/\delta$ is positive, the existence of a $\mu$-gradient acts 
as a limiting factor in the definition of the extension of the convective zone. If the layers close to the border of the convective zone do not fulfill the Schwartszschild criterion
$\nabla_T<\nabla_{ad}$, the borders of a convective zone defined according to the Ledoux criterion are secularly unstable, as a slow partial mixing through them produce the 
reduction of the local $\nabla_\mu$ value, thus allowing convection to encompass the borders themselves. 

However, at the \nane URCA shell the situation becomes more complex for two reasons: i) the $n$-rich isotope \isotope{23}{Ne} is locally converted into $e$-rich isotope 
\isotope{23}{Na}, thus producing a rapid decrease of the $\mu$ local value; ii) the energy released by URCA processes there is negative, as the $\varepsilon_M$ and 
$\varepsilon_K$ terms cancel out each others so that the energy sink $\varepsilon_\nu$ due to neutrino/antineutrino emission is the only local contribution. 
Both these two occurrences act in blocking the growth of the convective core, as the former tends to restore the local $\nabla_\mu$ and the latter determines a local negative 
value of luminosity and, hence, a negative temperature gradient $\nabla_T$. 

In order to investigate such an issue, we computed the model LED, having the same setup as model REF, but using the Ledoux criterion to define the border of zones unstable 
for convection. In the layers close to the convective zone, where it results $\nabla_L<\nabla_T<\nabla_{ad}$ (semiconvective layers), we assume that the local convective 
velocity is equal to 
\begin{equation}\label{e:vos}
v_{OS}=\alpha_{OS}\cdot v_{c}
\end{equation} 
where $v_{c}$ is the value of the convective velocity computed via MLT according to the Schwartzschild criterion and $\alpha_{OS}$ is a parameter whose value has been 
varied to study the depenendence of the obtained results on the mixing efficiency in the semiconvective zones. Once $v_{OS}$ has been determined, we recompute consistently 
the value of $\nabla_T$. 
We want to remark that, on a computational point of view, the determination \textit{a priori} of the real extension of the semiconvective region requires an iterative procedure. In fact, 
usually a molecular weight gradient occurs only at the border of the convective unstable region. As a consequence, partial convective mixing in progressively larger zones has to be 
tested to individuate the real extension of a semiconvective zone. This was first pointed out by \citet{castellani1971}, and successively remarked also by \citet{langer1983}. 
\begin{figure}
\plotone{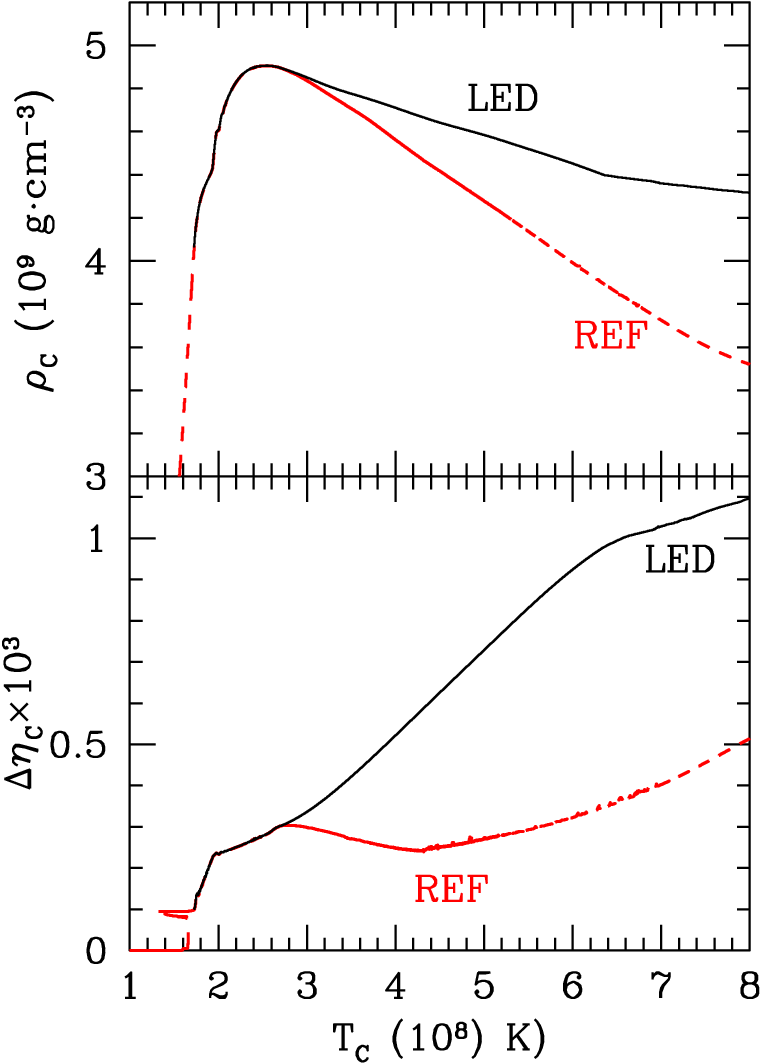}
\caption{Evolution of the density (upper panel) and of the neutronization variation (lower panel) as a function of the central temperature for model LED (black solid 
lines). For comparison model REF, is also displayed (long-dashed heavy red lines).\label{f:ledouxmod}}
\end{figure}

Our results, obtained by assuming $\mathrm{\alpha_{OS}=10^{-5}}$ are displayed in Figure~\ref{f:ledouxmod} where we report the same quantities as in 
Figure~\ref{f:ibenmod}, but for model LED as compared to model REF. The evolution of 
the two models is identical up to when the convective core approches the \nane URCA shell. At that epoch, due to contemporary action of weak processes determining the local 
cooling and, hence, the decrease of the local luminosity, and the local decrease of molecular weight due to the conversion of \isotope{23}{Ne} into \isotope{23}{Na}, 
we find that in the model LED implementing the Ledoux criterion the convective core stops at the \nane URCA shell. This determines a change in the slope $d\log T/d\log P$, 
causing a more rapid increase of the central temperature as compared to model REF. During the further evolution the mass of the convective core remains constant 
($\mathrm{M_{CC}}\simeq$ 0.449\msun); this limits the circulation of $e$-rich matter downward and, hence, the $\varepsilon_K$ energy contribution due to \ecap\, on \isotope{23}{Na}. As a 
consequence, a larger amount of carbon is consumed at the center to drive the accreting WD at the dynamical breakout. This determines a large increase of neutronization at the 
center and in the whole convective core. The further evolution does not present any particular difference, up to when the heating timescale at the center becomes shorter than 
that for \ecap\, on \isotope{23}{Na} at the center; hence, the increase of neutronization is mainly produced by the conversion of \isotope{13}{N} into \isotope{13}{C} (see 
the change in the slope of the curves in both panels in Figure~\ref{f:ledouxmod} at $\mathrm{T_C\simeq 6.4\times 10^8}$ K). 
We find that at the dynamical breakout, model LED has 
$\mathrm{\rho_{exp}=4.32\times 10^9}$\gcc, 
$\mathrm{M_{conv}^{max}=0.449}$\msun,
$\mathrm{\Delta M({^{12}C})=6.36\times 10^{-3}}$,
$\mathrm{\Delta\eta_C=1.097\times 10^{-3}}$, a factor 4.4 larger with respect to model REF.

We repeat the computation of model LED by changing the value of the $\mathrm{\alpha_{OS}}$. We find that the obtained results do not change for values in the range 
$\mathrm{10^{-7}\le\alpha_{OS}\le 1}$, clearly indicating that the local cooling at the \nane URCA shell determined by neutrino/antineutrino emission prevents 
any upward motion of the convective eddies, thus stopping the growth in mass of the convective core. We compute additional models by applying an overshooting at the border of 
the convective core. We consider an exponential decreasing profile of either the convective velocity or the Eulerian diffusion coefficient and we vary the length-scale adopted in the 
computation. In all the considered cases we found that the convective core is unable to encompass the URCA shell.

The analysis performed above clearly suggested that both the inclusion of the cooling of the surround produced by $e$-rich matter carried dwonward by convection and the 
use of the Ledoux criterion to define the border of convective unstable regions produce the same results of stopping the growth in mass of the convective core at the mass 
coordinate where the \nane URCA shell is located. The consequence of this is a lower convective motion of $n$-poor isotopes downward, so that the $\varepsilon_K$ energy 
term is largely reduced. Additionally, the inclusion of the negative $\varepsilon_W$ energy term magnifies such an effect. So we conclude that the proper treatment of URCA 
processes \textit{by including all the relevant energy contributions as reported in Eq.~\ref{e:e_urca2} and by accounting for the chemical composition gradient produced by nuclear 
burning} has the effects of determining a global cooling of the accreting WD. 

\section{The effect of \isotope{21}{(F,Ne)} URCA isobars}\label{s:finalsetup}

As pointed out first by \citet{iben1978b} and then remarked in P2017, the URCA pair \isotope{21}{(F,Ne)} plays an important role in determining the evolution of accreting WDs during the simmering phase. 
The initial abundance of \isotope{21}{Ne} in the WD progenitor star represents about 0.02\% of the total metallicity, even if it is largely produced during the main sequence evolution via the
\isotope{20}{Ne}(p,$\gamma$)\isotope{21}{Na}($\beta^+$)\isotope{21}{Ne}. Later on, during the central He-burning phase \isotope{21}{Ne} is slightly reduced via the 
\isotope{21}{Ne}(p,$\gamma$)\isotope{22}{Na}. As a result, in the WD model at the beginning of the accretion phase the \isotope{21}{Ne} abundance is about one order of magnitude larger 
than the initial value, the exact scaling factor depending mainly on the mass of the progenitor and the initial metallicity. 
Later on, during the simmering phase \isotope{21}{Ne} is produced via \ncap\, on \isotope{20}{Ne}. 

\begin{figure}
\gridline{\fig{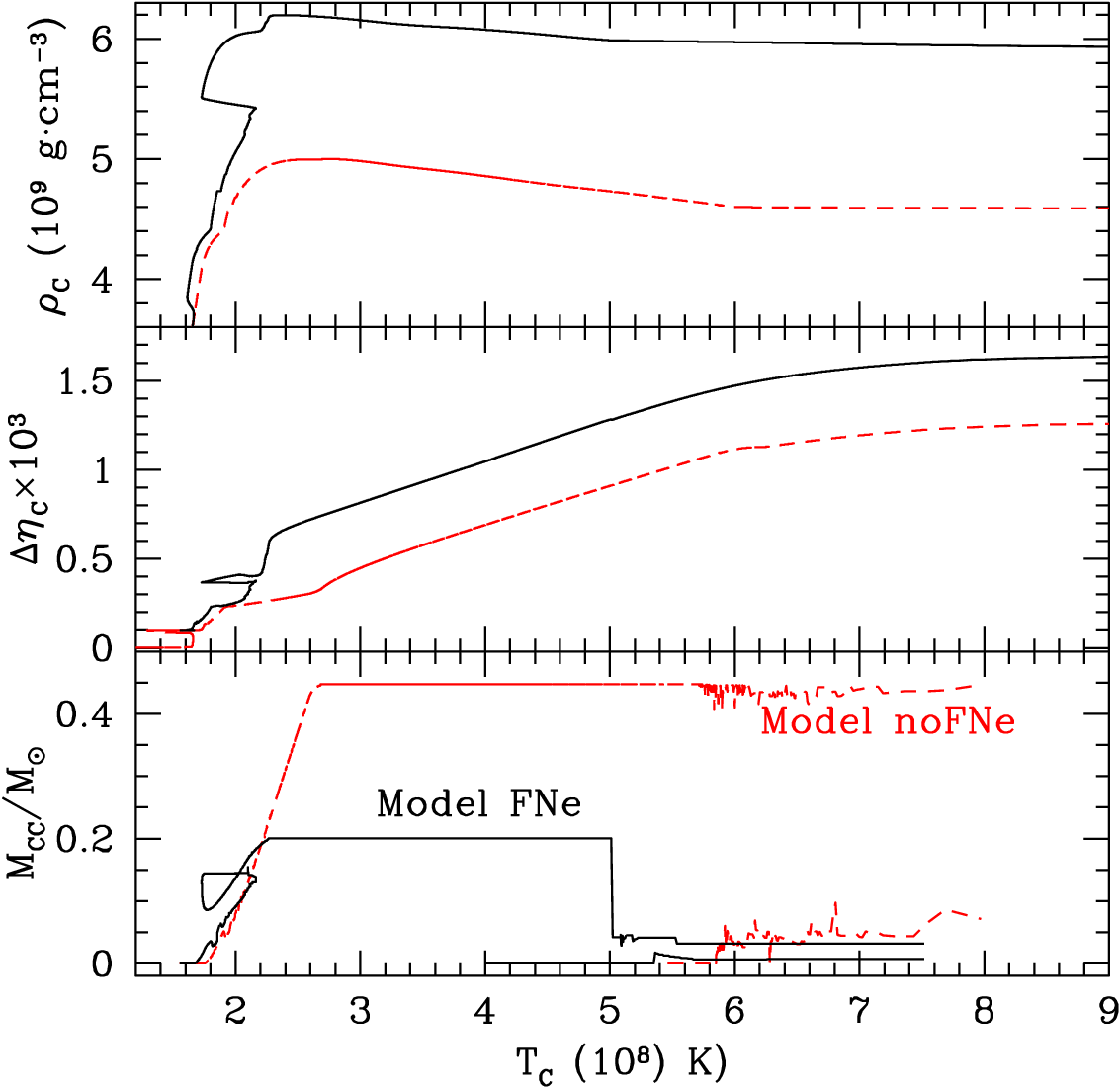}{\columnwidth}{}
            }
\caption{Evolution of the central density (upper panel) and of the central neutronization variation (middle panel) as a function of the central temperature for model FNe (black solid 
lines) and model noFNe (dashed red lines). In the lower panel we report the evolution of the border of the convective core.\label{f:FNeUrca}}
\end{figure}

In order to illustrate the effects of the \isotope{21}{(F,Ne)} URCA shell on the evolution of accreting WDs, we consider once again our test case, \textit{i.e.} the 
ZSUN WD model accreting CO rich matter at $10^{-7}$\msun\pyr, having an initial \isotope{21}{Ne} abundance equal to $3.74\times 10^{-5}$ by mass fraction. 
We include all the energy contributions related to weak processes as detailed in Eq.~\ref{e:e_urca2} and we define the borders of zone unstable for convection 
according to the Ledoux criterion. We adopt the full nuclear network detailed above and we compute two models, one including (model FNe) and another not 
including (model noFNe) weak processes involving \isotope{21}{F} and \isotope{21}{Ne}. 
The results are displayed in Figure~\ref{f:FNeUrca}, where we report the run of central density (upper panel), of neutronization variation at the center (middle panel) 
and of the mass extension of the convective core (lower panel) as a function of central temperature for these two models.

As the central density approaches the threshold value for \ecap\, on \isotope{21}{Ne}, ($\rho_{th}=3.78\times 10^9$\gcc), in model FNe the local temperature decreases 
due to emission of neutrino/antineutrino, the mass and kinetic energy terms in Eq.~\ref{e:e_urca2} canceling out each other. During this phase and up to when the central 
density does not exceed $\sim 5\times 10^9$\gcc, the \isotope{21}{Ne} nucleosynthesis is very scarce due to the fact that the equilibrium abundance of neutrons and of 
\isotope{17}{O} abundance is very low, so that neither $\alpha-$ nor $n-$channel production are efficient.
As a consequence, the evolution of model FNe is quite similar to that of model noFNe, even if it occurs at a slightly lower temperature: central convection sets in and when the 
central density exceeds the corresponding $\rho_{th}$ values, URCA triplets \isotope{28}{(Si,Al,Mg)} and \isotope{24}{(Mg,Na,Ne)} become active. 
However, when the central density exceeds $4.42\times 10^9$\gcc, the border of the convective core attains the location of the \isotope{21}{(F,Ne)} URCA shell. As a consequence,
the innermost zone of the core experience a thermal oscillation and the convective core mass slightly reduces. The same occurs also later on when the central density exceeds 
$\sim 4.7\times 10^9$\gcc. Such a behavior is caused by the interplay between the URCA triples \isotope{29}{(Si,Al,Mg)} and \isotope{24}{(Mg,Na,Ne)} with the 
\isotope{21}{(F,Ne)} URCA shell. As a net result it comes out that model FNe is not heated efficiently by the circulation of isobars involved in URCA processes, as it occurs 
in model noFNe. This produces the contraction of the innermost zones so that the model attains definitively larger central density. 
\begin{figure}
\gridline{\fig{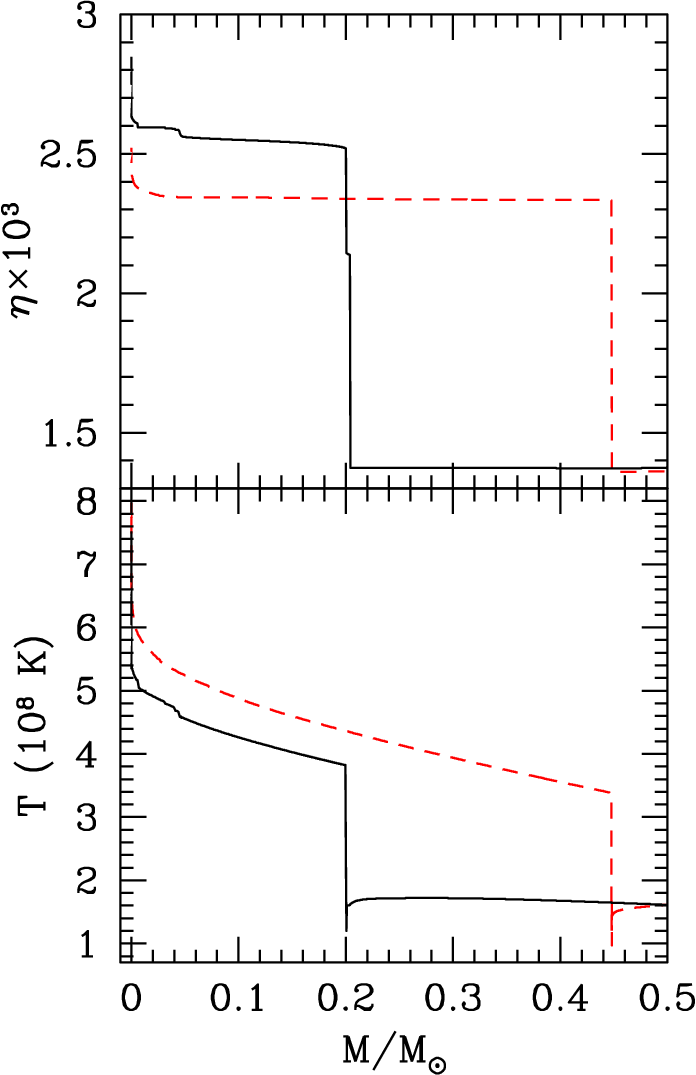}{\columnwidth}{}
            }
\caption{Neutronization (upper panel) and thermal profile (lower panel) as a function of the mass coordinate for the same models as in Figure~\ref{f:FNeUrca} at the epoch 
when the central temperature attains $8\times 10^8$ K.\label{f:FNeprofile}}
\end{figure}

When the central density exceeds $\sim 5.4\times 10^9$\gcc, model FNe experiences a huge thermal oscillation. At the beginning the circulation of $e-$rich material from the 
\isotope{21}{(F,Ne)} URCA shell inward produce the very efficient cooling of the center, while the border of the convective core remains close to the URCA shell, located at 
the mass coordinate $\sim 0.154$\msun. 
Later on, due to the decrease of the temperature in the convective core, the energy production via nuclear burning decreases, so that the convective core reduces. Hence, 
due the contraction of the innermost zones temperature starts to increase again, and the convective core grows in mass up to the \isotope{21}{(F,Ne)} URCA shell, which is 
now located at the mass coordinate $\sim 0.2$\msun. The occurrence of the large thermal oscillation is unavoidable. We tried to increase the spatial and temporal resolution, 
but the only result was to delay or anticipate its occurrence. We remark that the time and mass grid resolution adopted in the computation of this models represents the best 
choice to avoid spurious thermal oscillations 
whose only effects is to artificially increase the neutronization inside the convective core. In this regard, please note that the thermal oscillation described above and displayed 
in Figure~\ref{f:FNeUrca} has practically no effects on the neutronization at the center which is a tracer of the average neutronization in the convective core (see the middle 
panel). 

As a result, the inclusion of the \isotope{21}{(F,Ne)} URCA pairs determines a larger central density, a smaller convective core and a larger neutronization in the innermost zones. 
We want to remark that the ultimate consequence of the inclusion of this URCA pairs produces a lower total neutronization during the simmering phase in model FNe as 
compared to the noFNe one. In fact, in the latter the total variation of the neutronization along the whole CO WD $\Delta\eta_{tot}$ is $3.96\times 10^{-4}$ while in 
the former it is $2.40\times 10^{-4}$. Notwithstanding, at the explosion epoch, the average neutronization level in the innermost zones of the accreting WD is larger in model 
FNe because the maximum mass extension of the convective core is lower (see Figure~\ref{f:FNeprofile}). 

A further inspection of Figures \ref{f:FNeUrca} and \ref{f:FNeprofile} reveals that the use of the Ledoux criterion in defining the zones unstable for convection has an importan effect 
in determining the thermal and chemical stratification in the zones close to the center. In particular, as soon as the nuclear timescale at the center becomes comparable to that 
for energy transport via convection, the mixing efficiency of nuclearly processed matter from the center outward decreases. This produces a chemical composition gradient which 
acts as a stabilizing mechanism in the Ledoux criterion. As a consequence, the center of the accreting WD becomes radiative. In model FNe this determines also a decrease of the 
energy flux in the layers above, so that the convective core becomes a convective shell, having its inner border at $M\sim 0.007$\msun\, and extension $\sim 0.025$\msun. 
In model noFNe, the inner border of the convective shell is located at $\sim 0.08$\msun, even if the outer border is practically coincident with the maximum extension of the 
previous convective core. Such an occurrence determines the rapid increase of the temperature at the center, as clearly visible in both upper panel of 
Figure~\ref{f:FNeUrca} and lower panel Figure~\ref{f:FNeprofile}, respectively.

\begin{deluxetable*}{lrrrrr}
\tablecaption{Model inputs and results. \label{t:finalmodel}}
\tablehead{
\colhead{Model} & \colhead{Z14} & \colhead{Z63} & \colhead{Z12} & \colhead{Z22} & \colhead{Z42} 
}
\startdata
$\mathrm{Z_{ini}}$ ($\mathrm{10^{-3}}$)    & 0.245 & 6.000 & 13.80 & 20.00 & 40.00 \\ 
$\mathrm{X({^{21}Ne})\ ({10^{-5})}}$       & 0.139 & 2.230 & 3.740 & 3.080 & 4.020 \\ 
$\mathrm{M_{acc}}$ (\msun)                 & 0.570 & 0.568 & 0.567 & 0.565 & 0.560 \\ 
$\mathrm{t_{acc}}$ (10$^6$ yr)             & 5.695 & 5.679 & 5.665 & 5.652 & 5.605 \\ 
$\mathrm{t_{simm}}$ (10$^4$ yr)            & 6.294 & 5.504 & 5.853 & 6.335 & 8.518 \\ 
$\mathrm{\rho_{ign}}$ ($10^9$\gcc)         & 3.547 & 4.159 & 4.250 & 4.198 & 4.278 \\
$\mathrm{T_{ign}}$ ($10^8$ K)              & 1.941 & 1.858 & 1.693 & 1.643 & 1.475 \\
$\mathrm{\rho_{simm}}$ ($10^9$\gcc)        & 3.847 & 4.148 & 4.194 & 4.182 & 4.205 \\
$\mathrm{T_{simm}}$ ($10^8$ K)             & 2.021 & 1.854 & 1.693 & 1.631 & 1.450 \\
M(\isotope{12}{C})$_{ini}$ (\msun)         & 0.318 & 0.319 & 0.315 & 0.306 & 0.279 \\
M(\isotope{12}{C})$_{fin}$ (\msun)         & 0.316 & 0.317 & 0.313 & 0.304 & 0.275 \\
$\Delta$M(\isotope{12}{C})($10^{-3}$\msun) & 1.725 & 2.069 & 2.089 &2.248 & 3.455 \\
$\mathrm{M^{max}_{conv}}$ (\msun)          & 0.115 & 0.138 & 0.152 & 0.166 & 0.237 \\
$\mathrm{\rho_{exp}}$ ($10^9$\gcc)         & 5.022 & 5.273 & 5.447 & 5.504 & 6.344 \\
$\mathrm{\eta_{exp} (10^{-3})}$            & 1.200 & 1.913 & 2.738 & 3.495 & 5.526 \\
$\mathrm{\overline{\eta}_{exp} (10^{-3})}$ & 0.942 & 1.637 & 2.430 & 3.193 & 4.606 \\
$\mathrm{\eta_{c,0} (10^{-3})}$            & 0.022 & 0.570 & 1.278 & 1.883 & 3.690 \\
\hline                                                                              
\enddata
\tablecomments{
${\rm X({^{21}{Ne}})}$: Mass fraction abundance of \isotope{21}{Ne} in the initial WD model; 
${\rm M_{acc}}$: total accreted mass; 
${\rm t_{acc}}$: time from the onset of mass transfer up to the explosion; 
${\rm t_{simm}}$: time from the onset of convection up to the explosion; 
${\rm\rho_{ign}}$ and ${\rm T_{ign}}$: central density  and temperature at C-ignition; 
${\rm\rho_{simm}}$ and ${\rm T_{simm}}$: 
central density and temperature at the onset of convection; 
M(\isotope{12}{C})$_{ini}$ and M(\isotope{12}{C})$_{fin}$:  amount of \isotope{12}{C} in the innermost 0.8\msun\ 
of the initial WD and at the explosion; 
$\Delta$M(\isotope{12}{C}): amount of \isotope{12}{C} consumed via nuclear burning up to the explosion in the innermost 0.8\msun\ zone;
${\rm M^{max}_{conv}}$: maximum extension of the convective zone; 
$\rho_{\rm exp}$, $\eta_{\rm exp}$: density and neutronization values at the center, where the explosion occurs; 
${\overline\eta}_{\rm exp}$: neutronization at the explosion averaged over the convective zone; 
$\eta_{\rm c,0}$: initial value of central neutronization.}
\end{deluxetable*}

\section{Accreting WDs with different initial Z}\label{s:newmodel}

\begin{figure}
\gridline{\fig{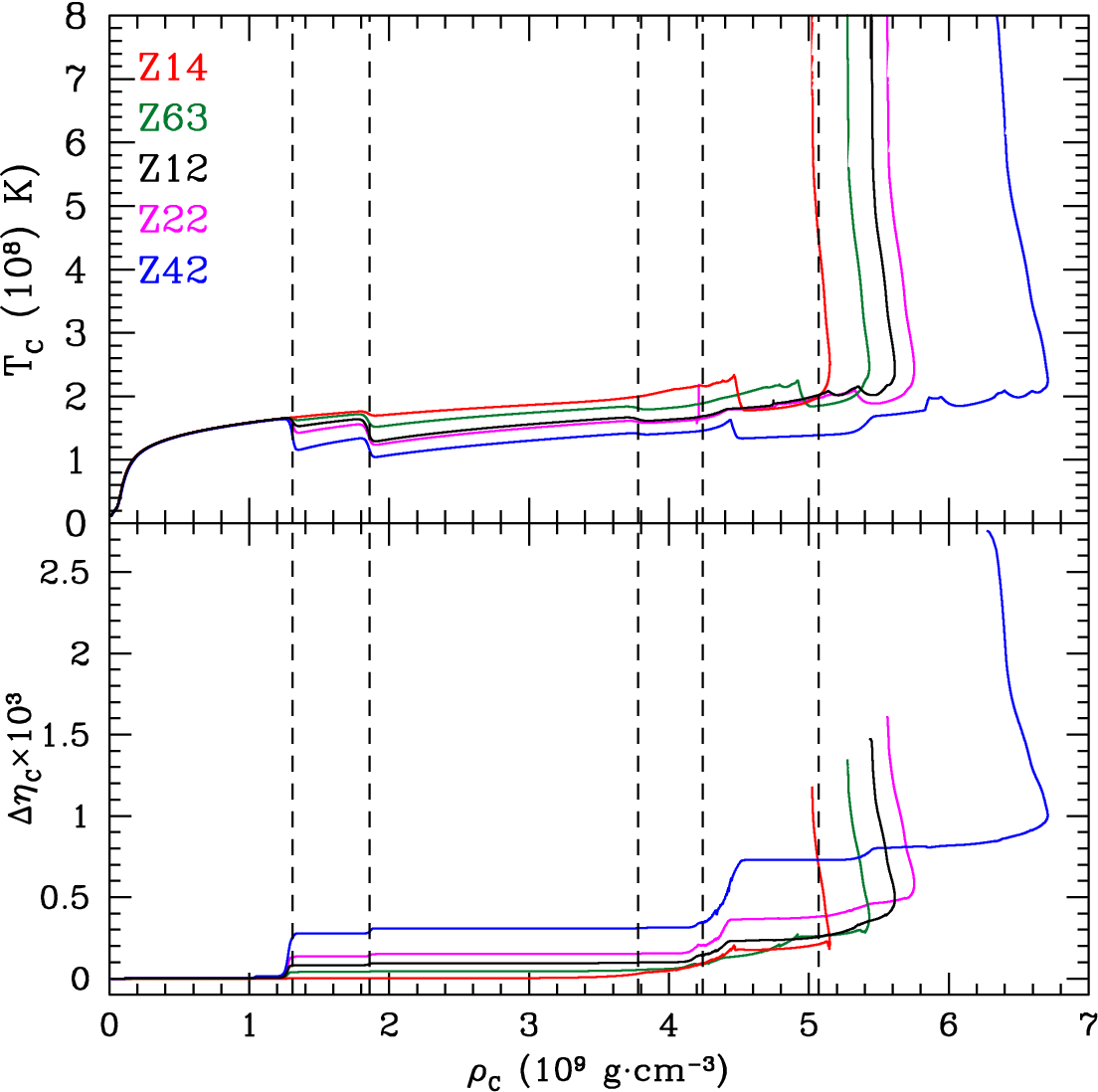}{\columnwidth}{}
            }
\caption{Evolution of the central temperature $T_C$ (in $10^8$ K - upper panel) and of variation of neutronization at the center $\Delta\eta_C$ (in $10^{-3}$ units - lower panel) as a function of the 
central density for all the models listed in Tab.~\ref{t:finalmodel}. Vertical dashed lines from left to right mark the values of threshold density for \ecap\ on \isotope{25}{Mg}, 
\isotope{23}{Na}, \isotope{21}{Ne}, \isotope{28}{Si}, and \isotope{24}{Na}, respectively. Please, note that the line corresponding to the $\rho_{th}$ value for \ecap\ on \isotope{24}{Mg} is 
indistinguishable in the plot from that of \isotope{28}{Si}. \label{f:finalphys}}
\end{figure}

In order to investigate the effects of different initial chemical composition of the WD progenitors on the simmering phase, we adopt five different starting WD models, with total metallicity 
of the progenitor equals to Z=$\mathrm{(0.0245,\ 0.6,\ 1.38,\ 2\ and\ 4)\times 10^{-2}}$ and labeled in the following as Z14, Z63, Z12, Z22 and Z42, respectively. Please, 
note that models Z14, Z12 and Z42 are the same as ZLOW, ZSUN and ZHIGH in P2017, respectively. We accrete CO rich matter according to the same procedure described in P2017 at a rate 
of \mdot=$10^{-7}$\msun\pyr. In the computation 
we adopt the same nuclear network as for model FNe above, while, as in P2017, we adopt the advective scheme for the convecting mixing. We define zones unstable for convection according to the 
Ledoux criterion and we include in the energy conservation equation all the contributions related to weak processes according to Eq.~\ref{e:e_urca2}.
Our results are summarized in Table~\ref{t:finalmodel} where we report several physical and chemical 
properties characterizing the simmering phase of the considered models. In Figure~\ref{f:finalphys} we report the evolution of the center in the $\rho-T$  (upper panel) and the value at the center 
of neutronization as a function of the central density.
Figure~\ref{f:finalphys} displays the onset of various URCA shell and their effect on the thermal evolution of the accreting WD. In particular, as it is well known, during the accretion phase 
\ecap s on \isotope{25}{Mg}, \isotope{23}{Na} and \isotope{21}{Ne} determines a cooling of the innermost zones of the accreting WDs and, at the same time, an increase of the local 
value of neutronization. The effect of the \isotope{21}{(Ne,F)} URCA shell is completely negligible in model Z14, as the \isotope{21}{Ne} abundance in the initial WD is very low (see Table~\ref{t:finalmodel}). 
In this model the central temperature continues to increase due to the homologous compression of the whole WD as determined by the continuous mass deposition. 
Moreover, Figure~\ref{f:finalphys} also discloses that the activation of \ecap\ on \isotope{24}{Mg} determines a rapid heating of the center, so that convection sets in (see the values 
of $\rho_{sim}$ in Table~\ref{t:finalmodel}). Also in this case, the evolution of model Z14 is different, as the onset of convection occurs at lower density. 

At this phase of the evolution, C-burning is fully active, so that \isotope{23}{Na} and \isotope{21}{Ne}\footnote{We recall that \isotope{21}{Ne} is produced via \ncap\ on \isotope{20}{Ne}, the 
neutrons being produced by the reaction chain \isotope{12}{C}(p,$\gamma$)\isotope{13}{N}($\beta^+$)\isotope{13}{C}($\alpha$,n)\isotope{16}{O}.} are produced. When central convection attains 
the \isotope{21}{(Ne,F)} URCA shell, the inward motion of e-rich matter, the cooling produced by the $\nu{\overline{\nu}}$ at the URCA shell, and the existence of a large $\mu$-gradient determined by 
the nuclear evolution in the innermost zone of the accreting WD stop the growth of the convective core and the innermost zones cool down. The following evolution is characterized by a reduction of the 
mass of the convective core, while the continuous mass deposition determines the compressional heating of the center and the increase of the whole density profile. As a matter of fact, the mass 
location of the \isotope{21}{(Ne,F)} URCA shell moves outward. In this phase nuclear burning occurs at a reduced rate, so that neutronization at the center is almost constant. Later on, 
when nuclear burning fully resume, the convective core start to increase once again, diluting the $n$-rich material and, hence, limiting the increase of $\eta_C$. When convection attains the \isotope{21}{(Ne,F)}
URCA shell, the local $\mu$ gradient stops the transport of thermal energy and prevents any further mixing of neutronized material outward. From this moment up to the explosion the external 
border of the convective unstable zone does not change anymore. 

Models with different initial metallicity experience different cooling during the pre-simmering phase, and this determines a different location of the \isotope{21}{(Ne,F)} URCA shell (the lower the initial Z, the more internal 
this URCA shell). As a consequence the maximum extension of the convective core depends also on the initial metallicity (see Table~\ref{t:finalmodel}). The occurrence that the convective core is limited by the 
URCA shell limits the circulation of neutron poor material so that such an heating mechanism has a very low efficiency. As a consequence the evolution up to the explosion is driven by C-burning, which, in turn, 
determines a large increase of the neutronization level in the innermost region (for instance, compare the run of $\eta_C$ in Figure~\ref{f:finalphys} with those in Figure~\ref{f:new_urca}). 
On the other hand, the fact that the convective core is small determines a rapid increase of the temperature in the burning region. 
\begin{figure}
\plotone{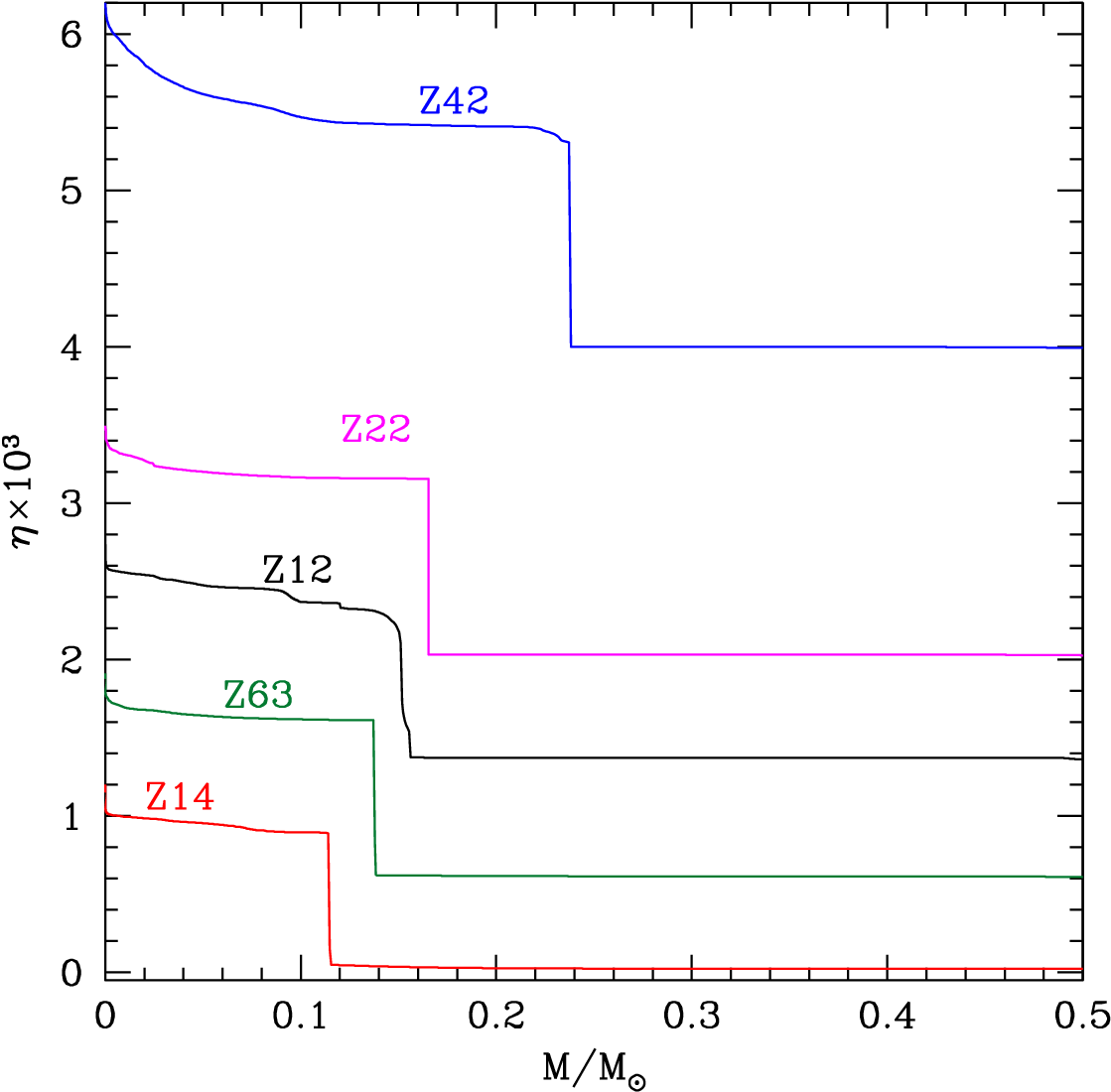}
\caption{Neutronization as a function of the mass coordinate at the explosion epoch, for all the computed models. \label{f:fin_str}}
\end{figure}

In Figure~\ref{f:fin_str} we report the run of neutronization as a function of the mass coordinate for all the computed models at the epoch of explosion. \par

\begin{deluxetable*}{lrrrrr|rr|rr}
\tablecaption{The same quantities as in Table~\ref{t:finalmodel} but for models with different accretion rates, as reported in the first line of the table. \label{t:variousmdot}}
\tablehead{
\colhead{Model} & \colhead{R5m8} & \colhead{Z12} & \colhead{R2m7} & \colhead{R6m7} & \colhead{R9m7} & \colhead{C1m7} & \colhead{C9m7} & \colhead{M1m7} & \colhead{M9m7} 
}
\startdata
$\mathrm{\dot{M}}$ ($\mathrm{10^{-8} M_\odot\cdot yr^{-1}}$)  & 5.0   & 10.0  & 20.0  & 60.0  & 90.0  & 10.0  & 90.0  & 10.0  & 90.0  \\ 
$\mathrm{M_{acc}}$ (\msun)                                    & 0.568 & 0.567 & 0.566 & 0.561 & 0.562 & 0.567 & 0.566 & 0.384 & 0.388 \\ 
$\mathrm{t_{acc}}$ (10$^6$ yr)                                & 11.35 & 5.665 & 2.828 & 0.935 & 0.625 & 5.668 & 0.629 & 3.839 & 0.431 \\ 
$\mathrm{t_{simm}}$ (10$^4$ yr)                               & 14.71 & 5.853 & 3.171 & 0.221 & 0.038 & 6.097 & 0.027 & 6.187 & 0.963 \\ 
$\mathrm{\rho_{ign}}$ ($10^9$\gcc)                            & 4.298 & 4.250 & 3.625 & 1.500 & 1.236 & 4.239 & 1.236 & 4.272 & 4.442 \\
$\mathrm{T_{ign}}$ ($10^8$ K)                                 & 1.463 & 1.693 & 1.901 & 2.059 & 2.153 & 1.687 & 1.933 & 1.678 & 1.378 \\
$\mathrm{M_{sim}^{in}}$ ($10^{-2}$\msun)                      & 0.000 & 0.000 & 0.000 & 0.153 & 2.365 & 0.000 & 14.55 & 0.000 & 0.000 \\
$\mathrm{\rho_{simm}}$ ($10^9$\gcc)                           & 4.270 & 4.194 & 1.872 & 3.743 & 3.519 & 4.193 & 3.471 & 4.200 & 4.287 \\
$\mathrm{T_{simm}}$ ($10^8$ K)                                & 1.454 & 1.693 & 1.526 & 2.394 & 2.551 & 1.668 & 2.628 & 1.652 & 1.310 \\
M(\isotope{12}{C})$_{fin}$ (\msun)                            & 0.313 & 0.313 & 0.314 & 0.311 & 0.312 & 0.314 & 0.310 & 0.313 & 0.312 \\
$\Delta$M(\isotope{12}{C})($10^{-3}$\msun)                    & 2.512 & 2.089 & 1.700 & 4.583 & 3.553 & 1.811 & 5.104 & 2.488 & 3.067 \\
$\mathrm{M^{max}_{conv}}$ (\msun)                             & 0.202 & 0.152 & 0.109 & 0.323 & 0.347 & 0.162 & 0.446 & 0.167 & 0.235 \\
$\mathrm{\rho_{exp}}$ ($10^9$\gcc)                            & 6.007 & 5.447 & 5.029 & 3.668 & 3.683 & 5.564 & 3.479 & 5.532 & 6.315 \\
$\mathrm{\eta_{exp} (10^{-3})}$                               & 2.726 & 2.738 & 2.767 & 2.754 & 2.320 & 2.753 & 2.352 & 2.820 & 3.182 \\
$\mathrm{\overline{\eta}_{exp} (10^{-3})}$                    & 2.355 & 2.430 & 2.526 & 2.188 & 1.952 & 2.240 & 2.118 & 2.503 & 2.326 \\
$\mathrm{M_{exp}}$ ($\mathrm{10^{-3}}$\msun)                  & 0.000 & 0.000 & 0.000 & 0.000 & 9.078 & 0.000 & 91.20 & 0.000 & 0.000 \\
\hline                                                                              
\enddata
\tablecomments{$\mathrm{M_{sim}^{in}}$ represents the mass coordinate of the inner border of the convective zone at the onset 
of the simmering phase; 
$\mathrm{M_{exp}}$ is the mass coordinate where the thermonuclear runaway occurs.}
\end{deluxetable*}

\section{Effects of different \mdot, \mwd\, and cooling age} \label{s:variation}

As it is well known, for CO WDs attaining \mch\, via mass accretion, the physical conditions at the onset of C-burning and, hence, at the explosion depend on the relative ratio of the compressional heating 
and of the thermal diffusion timescales. The former depends on the value of the mass transfer rate, while the latter is fixed by the thermal content of the accreting WDs, as determined by the cooling age before the 
onset of mass transfer, and on the WD initial mass. In order to investigate how the properties during the simmering phase depends on \mdot, \mwd\, and on the cooling age, we compute three additional sets of models. 
In \textit{Set 1} we adopt as initial model the 0.8 \msun\, CO WD with Z=1.38$\times 10^{-2}$ and we compute evolutionary sequence by adopting \mdot=0.5, 2, 6 and 9$\times 10^{-7}$\msun\pyr 
(models R5m8, R2m7, R6m7 and R9m7, respectively). 
In \textit{Set 2} we let the same CO WD model of \textit{Set 1} to cool down for additional $t_{cool}\simeq $1Gyr, up to when its central temperature attains $T_C=7.9\times 10^{7}$K; then we compute two 
evolutionary sequence by adopting \mdot=1 and 9$\times 10^{-7}$\msun\pyr\, (models C1m7 and C9m7, respectively). 
In \textit{Set 3} we accrete matter at \mdot=$10^{-7}$\msun\pyr on the same initial CO WD of \textit{Set 1}. When the total mass of the accreting WD is equal to 1\msun, we stop the accretion and let the model 
to cool down up to when its central temperature attains the value $T_C=1.13\times 10^7$ K. Hence, we compute two models with the same accretion rates as in \textit{Set 2} (models M1m7 and M9m7, respectively). 
In Table~\ref{t:variousmdot} we report selected physical and chemical quantities characterizing the accretion and the simmering phases for models in \textit{Set 1, 2} and \textit{3}. 

\begin{figure}
\gridline{\fig{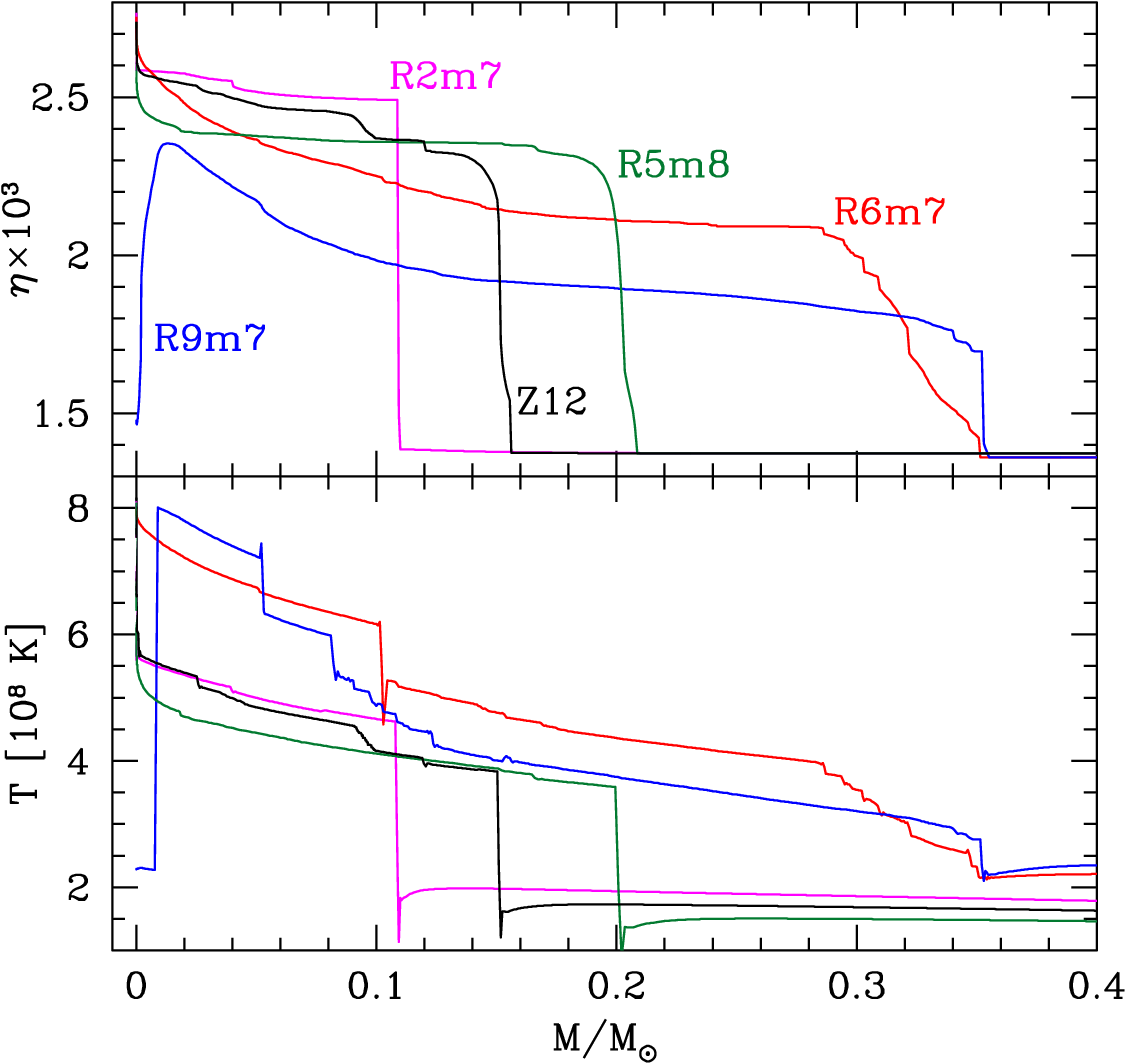}{\columnwidth}{}
            }
\caption{Neutronization (upper panel) and temperature (lower panel) profiles as a function of the mass coordinate at the explosion epoch, for the ZSUN WD model accreting matter at different accretion rate. \label{f:variousmdot}}
\end{figure}

Concerning the effects of different \mdot\, (models in \textit{Set 1}), 
the behaviour as well as the physical and chemical properties at the explosion of models with \mdot$\le 2\times 10^{-7}$\msun\pyr\, are easily explained when considering that, decreasing the accretion 
rate, the amount of mass to be accreted to trigger the explosion is larger. As a consequence, the lower the accretion rate, the higher the density level along the whole accreting structure and, hence, 
the more external the maximum mass coordinate attained by the \isotope{21}{(F,Ne)} URCA shell. This implies that for lower accretion rates the mass of the convective core is larger and, hence, a 
greater amount of \isotope{12}{C} has to be burnt to produce the explosion. However, even if in the considered range of \mdot the convective core almost doubles and the consumed carbon increases by 
$\sim$35\%, the average neutronization in the convective core decreases by only $\sim$7\% while the neutronization value at the center is practically unaffected (see also Fig. \ref{f:variousmdot}). 
The situation for model R6m7 with \mdot=$6\times 10^{-7}$\msun\pyr\, is quite different as C-burning is ignited close to the epoch of the \ecap\, on \isotope{21}{Ne} activation at the center.
This implies that C-ignition occurs off-center, due to the interplay among the compressional heating, of the inward diffusion of thermal energy from the accreted layers, and of the local cooling due to electron capture at the center. 
As a consequence, at the onset of the simmering phase a convective shell 
forms at the mass coordinate $M_{sim}^{in}\simeq 0.015$\msun, while the central region is cooled down by the conversion of \isotope{21}{Ne} into \isotope{21}{F}. However, as the convective shell grows in mass 
outward, the innermost zones start to expand so that the density at center decreases and the energy contribution from the \ecap\, on \isotope{21}{Ne} decreases. Contemporary thermal energy flows from the convective shell inward, 
heating up rapidly the center so that the thermonuclear runaway occurs there. As the \isotope{21}{(F,Ne)} URCA shell is ineffective in this model the resulting convective core can grows in mass up to the \isotope{23}{(Ne,Na)} 
URCA shell, so that the amount of carbon to be consumed to trigger the explosion results definitively larger than that for model R2m7 (see Table~\ref{t:variousmdot}). 
Model R9m7 has the same behaviour even if the compressional heating timescale is definitively shorter than the inward thermal diffusion timescale, thus determining the occurrence of C-burning very far from center. As in model 
R6m7 at the onset of the simmering phase a convective shell forms, but now its inner border is more external and it is located at $\sim$0.024\msun. This prevents any efficient heating of the central region of the accreting WD so 
that the thermonuclear runaway occurs off center. The other important difference between models R6m7 and R9m7 is that the latter has a larger thermal content at the onset of the simmering phase, so that a smaller amount 
of \isotope{12}{C} has to be burnt to trigger the explosion, even if the maximum extension of the zone unstable for convection is larger.
By considering the whole set of models with different accretion rates  it comes out that the average neutronization in the innermost zone of the WD varies by no more than 13\%.

By comparing models C1m7 and Z12 ones, it comes out that the cooling age, \textit{i.e.} the thermal content of the initial CO WD, has a negligible effects for low values of the accretion rate, 
because, in this case, the heating of the innermost zones of the star is driven by the homologous compression only. On the other hand, for large values of \mdot, the comparison of models R9m7 and C9m7 
demonstrates that the lower the initial thermal content of the accreting WD the more external the mass coordinate where C-burning occurs and, hence, the more external the point where the explosion occurs. 
In addition, while the maximum extension of the convective core is quite similar, the amount of \isotope{12}{C} consumed in model C9m7 is 30\% larger than in models R9m7, because a larger amount of nuclear energy 
has to be injected to heats up the accreting WD and to drive it up to the explosion. As a consequence, in models C9m7 the average neutronization level inside the convective shell is $\sim$8\% larger than in model R9m7, 
even if the $\eta$ value at the explosion point is largely unaffected. 

When considering models in \textit{Set 3}, it comes out once again that for low values of the accretion rate the physical and chemical properties at the C-ignition epoch and during the simmering phase up to the explosion 
are largely independent on the initial mass of the accreting CO WDs. On the other hand, the thermal diffusion timescale in model M9m7 is definitively larger than the compressional heating one, so that the evolution of the innermost  
$\simeq$ 0.5\msun is driven by the homologous compression only. As a consequence, C-ignition occurs at the center for very large values of density and, as a whole, its evolution resembles quite closely that of 
models with lower CO WD initial mass and very low mass accretion rate. 

\section{Final considerations} \label{s:conclusion}
In this work we analyze URCA processes and their interplay with convection during the simmering phase of SNe Ia progenitors. By including all the relevant physical 
processes we determine accurately the physical and chemical profile of an accreting WD as well as the neutronization level at the explosion epoch. Our results 
suggest that:
\begin{enumerate}
\item complete nuclear network, at least up to \isotope{32}{S} should be used. In fact, by neglecting \ncap,\ \acap,\ and \pcap, the amount of \isotope{12}{C} consumed 
during the simmering is overestimated and, consequently, the final neutronization level results larger; 
\item all the relevant equation describing the physical and chemical structure of an accreting WD as well as its time evolution should be solved contemporary. In fact, 
by using the operator splitting technique, the effective energy injected during the simmering is overestimated so that the 
\isotope{12}{C} consumption results larger and also the neutronization level is erroneously overestimated;
\item the Ledoux criterion should be used in order to account for the existence of a $\mu$-gradient at a given URCA shell. Our results show that according to this 
criterion the convective core stops at the innermost URCA shell (\textit{i.e.} the \isotope{21}{(Ne,F)} one), thus reducing the inward circulation of $e$-rich isobars 
and the corresponding contribution to the total energy budget; 
\item the inward convective motion of $e$-rich isobar from an URCA shell determines an effective cooling and, in addition, prevents convection 
to encompass the URCA shell itself. Such an occurrence determines a larger C-consumption and, hence, a larger neutronization level and a higher density at the explosion epoch;
\item the chemical and physical structure at the explosion epoch depends on the assumed mixing efficiency, as determined by different numerical algorithms used to 
model convective mixing. 
\end{enumerate}
 
An inspection to  Table~\ref{t:finalmodel} and of Figures~\ref{f:finalphys} and \ref{f:fin_str}, clearly suggests that the properties of the accreting WDs during the simmering 
phase and at the explosion epoch depend on the initial metal content of the progenitor. This can be easily understood, because the larger the initial metallicity:
\begin{itemize}
\item the larger the abundance of URCA mother nuclei in the accreting WD and, hence, the stronger their cooling effects during the accretion phase;
\item the larger the amount of mass to be accreted to trigger the thermonuclear runaway;
\item the more external the position of the \isotope{21}{(F,Ne)} URCA shell and, hence, the larger the zone unstable for convection during the simmering;
\item the larger the amount of nuclear energy to be injected to trigger the explosion;
\item the larger the carbon consumed and, hence, the larger the neutronization level.
\end{itemize}
We want to remark once again that such a result is a direct consequence of the fact that the innermost URCA shell represents a barrier for convection so that circulation of 
URCA isobars across the shell is inhibited and their energy contribution to the heating up of the innermost zone is drastically reduced.  

For this reason, analogously to P2017, we define a final-to-initial neutron-excess relation (FINE) by interpolating the values of $\mathrm{\overline{\eta}_{exp}}$ as 
a function of $\mathrm{\eta_{c,0}}$, thus obtaining
\begin{eqnarray}
\overline{\eta}_{exp}&=&(1.080\pm 0.132)\times 10^{-3}+\\
& & (0.995\pm0.068)\times \eta_{c.0}\nonumber\\
R^2 &=& 0.9930 \nonumber
\label{e:fine}
\end{eqnarray}
On the other hand the value of $\mathrm{\eta_{c,0}}$ depends on the initial metallicity according to the relation:
\begin{eqnarray}
\eta_{c,0}&=& 1.213\times 10^{-5}+9.224\times 10^{-2}\times Z\\
R&=&1.000  \nonumber
\end{eqnarray}
so that Eq.~\ref{e:fine} becomes 
\begin{eqnarray}
{\overline{\eta}_{exp}} &=& (1.094\pm 0.143)\times 10^{-3} +\\
& & (9.168\pm 0.677)\times 10^{-2}\times Z\nonumber \\ 
R^2 &=& 0.9920 \nonumber
\end{eqnarray}

Such a relation can be used to infer the metallicity of stars progenitors of galactic SNe Ia, once a reliable estimate of the neutronization level at the explosion epoch 
is available. \cite{badenes2008} and \cite{park2013} derived the neutronization level of the Tycho and Kepler Supernova basing on the observed actual 
abundance of Mn and Cr, $\overline{\eta}_\mathrm{exp,Tycho}=4.36\times10^{-3}$ and $\overline{\eta}_\mathrm{exp,Kepler}=4.55\times10^{-3}$. 
By assuming that these values corresponds to the average neutronization of the two Supernovae at the explosion epoch, one can derive the progenitor metallicity to be:
$Z_\mathrm{Tycho}=0.0353=2.56Z_\odot$ and $Z_\mathrm{Kepler}=0.0373=2.71Z_\odot$

However, the estimation of the remnants metallicity in Tycho and Kepler assumed that the measured Mn and Cr were synthesized in the incomplete Si burning regime, which 
is expected to occur above the mass affected by convection in present models. As can be seen in Figure\ref{f:fin_str}, the neutron excess above the \isotope{21}{(F,Ne)} 
URCA shell is the same as in the progenitor WD before 
accretion begins. As a consequence, we recover the large values of Z derived by \citet{badenes2008,park2013} for the progenitors of these two historical SNe Ia.

Our results are in agreement with the findings by P2017, even if the physical and chemical properties of models at the explosion epoch are completely different. 
In fact, in P2017 not all the relevant energy contributions related to URCA processes are included consistently and the effects of $\mu$-gradient are not taken into 
account in defining the zones unstable for convection. As a consequence, in that work the amount of carbon consumed to trigger the explosion is about an order 
of magnitude larger than what estimated in the present work (compare the values of $\Delta$M(\isotope{12}{C}) reported in Table~\ref{t:finalmodel} and 
those in Tab. 2 in P2017). As a consequence, in P2017 the total neutronization produced during the evolution is also larger, even if the average value in the 
convective core is almost the equal to the one derived in the present work: this is a consequence of the fact that in P2017 almost the whole accreting WD is unstable for 
convection, while in our models the convective core extends only on the innermost zones. 

We have found that a large accretion rate is needed to affect significantly the evolution of the WD pre-explosion. Accretion rates as large as (6---9)$\times 10^{-7}$\msun\pyr\, imply a high rate 
of compressional heating, which shortens the time needed for carbon to achieve simmering conditions, first, and explosive conditions, later. The result is that the \isotope{21}{(F,Ne)} URCA pair influence 
is reduced, convection sets in a shell close to the center, the convective mass is larger (but far from the values close to $\sim$ 1\msun\, found in previous works), and the explosion density is smaller, 
on the order of $3.6\times 10^{9}$\gcc.
\begin{acknowledgments}
This publication is part of the project I+D+I  PGC2018-095317-B-C21 funded by MICIN/AEI/10.13039/501100011033  and FEDER “A way of doing Europe” (E.B. and I.D.)
L.P. and O.S. aknowledge finacial support from the INAF-mainstream project ``Type Ia Supernovae and their Parent Galaxies: Expected Results from LSST''.
O.S. and L.P. acknowledge their participation to the V:ANS project (Vanvitelli program on standard candles in astrophysics: Atomic and Nuclear physics in SNIa) supported by the Vanvitelli University.
\end{acknowledgments}

\appendix
\begin{deluxetable*}{lcccccccc}
\centering
\tablecaption{Computed models and adopted computational setup. \label{t:models}}
\tablehead{
\colhead{Model} & \colhead{Z} & \mwd (\msun) & \colhead{\mdot\ ($10^{-7}$\msun\pyr)} &\colhead{Nuc.Net.} & \colhead{Conv.Crit.} & \colhead{Mix.Sch.} & \colhead{Coupled} & \colhead{$\varepsilon_{URCA}$}
}
\startdata
K00  & $1.38\times 10^{-2}$ & 0.8 & 1.0 & SHORT & SCHWAR & DIF & YES  & Eq.~\ref{e:e_urca} \\
K28  & $1.38\times 10^{-2}$ & 0.8 & 1.0 & SHOR1 & SCHWAR & DIF & YES  & Eq.~\ref{e:e_urca} \\
K24  & $1.38\times 10^{-2}$ & 0.8 & 1.0 & SHOR2 & SCHWAR & DIF & YES  & Eq.~\ref{e:e_urca} \\
SEL  & $1.38\times 10^{-2}$ & 0.8 & 1.0 & SHORT & SCHWAR & DIF & NO   & Eq.~\ref{e:e_urca} \\
ADd  & $1.38\times 10^{-2}$ & 0.8 & 1.0 & SHORT & SCHWAR & ADV & NO   & Eq.~\ref{e:e_urca} \\
DFd  & $1.38\times 10^{-2}$ & 0.8 & 1.0 & SHORT & SCHWAR & DIF & NO   & Eq.~\ref{e:e_urca} \\
ADV  & $1.38\times 10^{-2}$ & 0.8 & 1.0 & SHORT & SCHWAR & ADV & YES  & Eq.~\ref{e:e_urca} \\
REF  & $1.38\times 10^{-2}$ & 0.8 & 1.0 & LONG  & SCHWAR & DIF & YES  & Eq.~\ref{e:e_urca} \\
W00  & $1.38\times 10^{-2}$ & 0.8 & 1.0 & LONG  & SCHWAR & DIF & YES  & Eq.~\ref{e:e_urca2} \\
W06  & $1.38\times 10^{-2}$ & 0.8 & 1.0 & LONG  & SCHWAR & DIF & YES  & Eq.~\ref{e:e_urca2} \\
WRK  & $1.38\times 10^{-2}$ & 0.8 & 1.0 & LONG  & SCHWAR & DIF & YES  & Eq.~\ref{e:e_urca2} \\
LED  & $1.38\times 10^{-2}$ & 0.8 & 1.0 & LONG  & LEDOUX & DIF & YES  & Eq.~\ref{e:e_urca} \\
noFNe& $1.38\times 10^{-2}$ & 0.8 & 1.0 & LONG  & LEDOUX & DIF & YES  & Eq.~\ref{e:e_urca2} \\
FNe  & $1.38\times 10^{-2}$ & 0.8 & 1.0 & FULL  & LEDOUX & DIF & YES  & Eq.~\ref{e:e_urca2} \\
Z14  & $2.45\times 10^{-4}$ & 0.8 & 1.0 & FULL  & LEDOUX & ADV & YES  & Eq.~\ref{e:e_urca2} \\
Z63  & $6.00\times 10^{-3}$ & 0.8 & 1.0 & FULL  & LEDOUX & ADV & YES  & Eq.~\ref{e:e_urca2} \\
Z12  & $1.38\times 10^{-2}$ & 0.8 & 1.0 & FULL  & LEDOUX & ADV & YES  & Eq.~\ref{e:e_urca2} \\
Z22  & $2.00\times 10^{-2}$ & 0.8 & 1.0 & FULL  & LEDOUX & ADV & YES  & Eq.~\ref{e:e_urca2} \\
Z42  & $4.00\times 10^{-2}$ & 0.8 & 1.0 & FULL  & LEDOUX & ADV & YES  & Eq.~\ref{e:e_urca2} \\
R5m8 & $1.38\times 10^{-2}$ & 0.8 & 0.5 & FULL  & LEDOUX & ADV & YES  & Eq.~\ref{e:e_urca2} \\
R2m7 & $1.38\times 10^{-2}$ & 0.8 & 2.0 & FULL  & LEDOUX & ADV & YES  & Eq.~\ref{e:e_urca2} \\
R6m7 & $1.38\times 10^{-2}$ & 0.8 & 6.0 & FULL  & LEDOUX & ADV & YES  & Eq.~\ref{e:e_urca2} \\
R9m7 & $1.38\times 10^{-2}$ & 0.8 & 9.0 & FULL  & LEDOUX & ADV & YES  & Eq.~\ref{e:e_urca2} \\
C1m7 & $1.38\times 10^{-2}$ & 0.8 & 1.0 & FULL  & LEDOUX & ADV & YES  & Eq.~\ref{e:e_urca2} \\
C9m7 & $1.38\times 10^{-2}$ & 0.8 & 9.0 & FULL  & LEDOUX & ADV & YES  & Eq.~\ref{e:e_urca2} \\
M1m7 & $1.38\times 10^{-2}$ & 1.0 & 1.0 & FULL  & LEDOUX & ADV & YES  & Eq.~\ref{e:e_urca2} \\
M9m7 & $1.38\times 10^{-2}$ & 1.0 & 9.0 & FULL  & LEDOUX & ADV & YES  & Eq.~\ref{e:e_urca2} \\
\hline                                                                              
\enddata
\end{deluxetable*}

\section{Summary of computed models}\label{a:computed}
In Table~\ref{t:models} we list all the models presented in the text and summarize the setup adopted in their computations. 
The meaning of the various columns is:
\begin{enumerate}
\item \textbf{Model}: label adopted in the text to address a given model; 
\item \textbf{$\mathrm{M_{WD}}$}: initial mass of the CO WD at the onset of the accretion phase;
\item \textbf{\mdot}: adopted accretion rate;
\item \textbf{Nuc.Net.}: Nuclear network adopted in the computation. The different possibilities are:\par\noindent
\textbf{SHORT}: same one as in \citet{piro2008}, but including also the \isotope{25}{(Mg,Na)} and \isotope{56}{(Fe,Cr,Mn)} URCA processes;\par\noindent
\textbf{SHOR1}: same as SHORT, but including also the \isotope{28}{(Si,Al,Mg)} URCA triplet (see \S\ref{s:preanalysis});\par\noindent
\textbf{SHOR2}: same as SHOR1, but including also the \isotope{24}{(Mg,Na,Ne)} URCA triplet (see \S\ref{s:preanalysis});\par\noindent
\textbf{LONG}: nuclear network including 52 isotopes all the strong and weak nuclear processes, except for weak processes on \isotope{21}{F} and \isotope{21}{Ne} (see \S\ref{s:finalsetup});\par\noindent
\textbf{FULL}: same as LONG, but including also the weak processes on \isotope{21}{F} and \isotope{21}{Ne} (see \S\ref{s:finalsetup});
\item \textbf{Conv.Crit.} adopted criterion to defined zones unstable for convection (see \S\ref{s:neutronization}):\par\noindent
\textbf{SCHWAR}: Schwartzshild criterion; \par\noindent
\textbf{LEDOUX}: Ledoux criterion;
\item \textbf{Mix.Sch.}: numerical scheme adopted to model convective mixing (see \S\ref{s:preanalysis}) :\par\noindent
\textbf{DIF}: diffusive scheme;\par\noindent
\textbf{ADV}: advective scheme\par\noindent
\textbf{SEL}: Sparks -Endal-Linearized scheme
\item \textbf{Coupled}: \textbf{YES}: all the equations describing the physical and chemical evolution of accreting WD are solved simultaneously; \textbf{NO}: the operator splitting technique is applied;
\item \textbf{$\varepsilon_{URCA}$}: energy contribution due to URCA processes and, more in general to weak processes, included in the equation for the conservation of energy.
\end{enumerate}

\bibliography{piersanti}{}
\bibliographystyle{aasjournal}

\listofchanges

\end{document}